\documentclass[twocolumn]{aastex631}

\usepackage{amsmath}
\usepackage{hyperref}
\usepackage{enumitem}
\usepackage{footnote}
\usepackage{tablefootnote}
\usepackage{threeparttable}

\usepackage{lipsum}
\usepackage{comment}
\usepackage{multirow}
\usepackage{enumitem}
\usepackage{hyperref}
\usepackage{changepage}

\usepackage{xcolor, soul}

\graphicspath{{./}}

\begin{document}

\title{Rapid Quenching of Galaxies at Cosmic Noon}

\author[0000-0002-8435-9402]{Minjung Park}
\affiliation{Center for Astrophysics $\mid$ Harvard $\&$ Smithsonian, 60 Garden St., Cambridge, MA 02138, USA}

\author[0000-0002-5615-6018]{Sirio Belli}
\affiliation{Dipartimento di Fisica e Astronomia, Università di Bologna,
Via Gobetti 93/2, 40129, Bologna, Italy}

\author{Charlie Conroy}
\affiliation{Center for Astrophysics $\mid$ Harvard $\&$ Smithsonian, 60 Garden St., Cambridge, MA 02138, USA}

\author{Sandro Tacchella}
\affiliation{Kavli Institute for Cosmology, University of Cambridge, Madingley Road, Cambridge, CB3 0HA, UK}
\affiliation{Cavendish Laboratory, University of Cambridge, 19 JJ Thomson Avenue, Cambridge, CB3 0HE, UK}

\author{Joel Leja}
\affiliation{Department of Astronomy $\&$ Astrophysics, The Pennsylvania State University, University Park, PA 16802, USA}
\affiliation{Institute for Computational $\&$ Data Sciences, The Pennsylvania State University, University Park, PA, USA}
\affiliation{Institute for Gravitation and the Cosmos, The Pennsylvania State University, University Park, PA 16802, USA}

\author[0000-0002-7031-2865]{Sam E. Cutler}
\affiliation{Department of Astronomy, University of Massachusetts, Amherst, MA 01003, USA}

\author{Benjamin D. Johnson}
\affiliation{Center for Astrophysics $\mid$ Harvard $\&$ Smithsonian, 60 Garden St., Cambridge, MA 02138, USA}

\author{Erica J. Nelson}
\affiliation{Department for Astrophysical and Planetary Science, University of Colorado, Boulder, CO 80309, USA}

\author{Razieh Emami}
\affiliation{Center for Astrophysics $\mid$ Harvard $\&$ Smithsonian, 60 Garden St., Cambridge, MA 02138, USA}

\begin{abstract}
The existence of massive quiescent galaxies at high redshift seems to require rapid quenching, but it is unclear whether all quiescent galaxies have gone through this phase and what physical mechanisms are involved.  To study rapid quenching, we use rest-frame colors to select 12 young quiescent galaxies at $z \sim 1.5$.  From spectral energy distribution fitting we find that they all experienced intense starbursts prior to rapid quenching. We confirm this with deep Magellan/FIRE spectroscopic observations for a subset of seven galaxies. Broad emission lines are detected for two galaxies, and are most likely caused by AGN activity. The other five galaxies do not show any emission features, suggesting that gas has already been removed or depleted. Most of the rapidly quenched galaxies are more compact than normal quiescent galaxies, providing evidence for a central starburst in the recent past. We estimate an average transition time of $300\,\rm Myr$ for the rapid quenching phase.  Approximately $4$\% of quiescent galaxies at $z=1.5$ have gone through rapid quenching; this fraction increases to $23$\% at $z=2.2$. We identify analogs in the TNG100 simulation and find that rapid quenching for these galaxies is driven by AGN, and for half of the cases, gas-rich major mergers seem to trigger the starburst. We conclude that these massive quiescent galaxies are not just rapidly quenched but also rapidly formed through a major starburst. We speculate that mergers drive gas inflow towards the central regions and grow supermassive black holes, leading to rapid quenching by AGN feedback.
\end{abstract}


\section{Introduction} \label{sec:intro}
Star-forming activity is one of the fundamental characteristics of galaxies and is closely related to various properties such as stellar mass, color, and morphology \citep[e.g.,][]{Strateva2001, Baldry2004, Wuyts2011, Bluck2014}. One of the most important unresolved questions in galaxy evolution is understanding how galaxies evolve from the star-forming to the quiescence phase. Several quenching mechanisms have been proposed, and recent studies have suggested two broad quenching processes with different timescales, namely ``rapid'' quenching and ``slow'' quenching \citep[e.g.,][]{Wu2018, Belli2019, Wild2020}. The post-starburst galaxies (PSB), originally known as E+A galaxies \citep{Dressler_Gunn_1983, Zabludoff1996}, i.e., elliptical but with an A-type young stellar spectrum, are thought to be in rapid transition from star-forming to quiescence. As the name suggests, they are thought to have had starbursts in the past but rapidly and recently quenched so that they are still dominated by young stars but without on-going star formation. Thus, these PSBs hold important clues about the rapid quenching processes \citep{French2021}.

The origin of the starburst is not yet clear. Gas-rich major mergers have been suggested as a possible scenario \citep[e.g.,][]{Barnes1991,Bekki2005,Snyder2011}. As galaxies undergo gas-rich mergers, gas can flow into the central region, triggering a starburst. However, it is not entirely clear whether mergers are always involved in this picture. Some studies have found evidence for merger-fueled central starbursts \citep[e.g.,][]{Puglisi2019}, while others suggest outside-in formation via dissipative collapse \citep[e.g.,][]{Tadaki2017, Tadaki2020}. At high redshifts, compaction processes driven by violent disk instability or misaligned gas streams could also trigger the central starburst \citep[e.g.,][]{Dekel2014WetNuggets, Zolotov2015CompactionNuggets, Tacchella2016b, Nelson2019_bulgediskz1}. The central starburst will then deplete the gas temporarily, suppressing star formation. Several studies have shown that galaxies could remain on the main sequence after significant central starbursts followed by gas compaction events \citep[e.g.,][]{Tacchella2016b, Cutler2022_differential_assembly_history}. Therefore, some other preventive mechanisms are required to make galaxies remain quiescent.

AGN activity seems to be an important mechanism for rapid quenching. It could both blow away gas (often referred to as kinetic feedback) and heat the surrounding medium and thus prevent cooling (thermal feedback). Many studies using simulations have shown that AGN activity is essential in reproducing post-starburst populations \citep[e.g.,][]{Pontzen2017, Davis2019, Zheng2020}. However, it is very challenging to directly observe evidence of the ongoing AGN activity. Several studies have shown that PSBs have emission diagnostics similar to LINERs \citep[e.g.,][]{French2015}. Galactic outflows have been observed for a number of post-starburst galaxies \citep[e.g.,][]{Baron2017, Maltby2019}, and given their high speed ($>1000\,\rm km\,s^{-1}$), the outflows are thought to be driven by ejective AGN feedback \citep[e.g.,][]{Forster-Schreiber2019}.

At high redshifts, it is often very challenging to spectroscopically confirm PSBs, thus, many studies focus on young quiescent galaxies (in many cases, photometrically selected) which are most likely to be recently and rapidly quenched \citep[e.g.,][]{Whitaker2012compact_PSB, Wild2016, Belli2019, Suess2020}. Therefore, it is not clear whether these young quiescent galaxies had a starburst in the past and then rapidly quenched (thus, truly ``post-starburst'') or simply had a sudden quenching after a relatively flat star formation histoy (SFH) \citep[e.g.,][]{Wild2020}. The fraction of quiescent galaxies that are young increases with redshift \citep{Whitaker2012, Wild2016, Belli2019}, suggesting that the rapid quenching process seems to become more important and common at high redshifts. Several studies have also attempted to constrain the quenching timescales of galaxies both in observations \citep[e.g.,][]{Tacchella2022_Fast_slow_quenching} and simulations \citep[e.g.,][]{Rodriguez-Gomez2019, Park2022}, both at low and high redshifts. They found that galaxies at high redshifts tend to be more rapidly quenched (typically $< 1\,\rm Gyr$), while at low redshifts, they have a broad range of quenching timescales (up to several Gyrs). The PSBs are also more frequently found at high redshifts ($>5\%$) than in the local universe ($<1\%$) \citep[e.g.,][]{Wild2016}, and the existence of quiescent galaxies found at very high redshifts ($z>3$) \citep[e.g.,][]{Franx2003, Forrest2020}, when the age of the Universe is less than a few Gyrs, requires a very rapid quenching process. Indeed, \cite{DEugenio2020InverseLEGA-C} studied 9 spectroscopically confirmed quiescent galaxies at $z\sim3$ and showed that their average spectra are very similar to those of PSBs.

However, it is not yet clear how much of the quiescent population was built up through the rapid quenching phase. Different studies have estimated how many quiescent galaxies have gone through the rapid quenching phase, and the conclusions are often sensitive to the definition of post-starburst and strongly depend on redshift. For example, \cite{Belli2019} identified rapidly quenched PSB galaxies as young quiescent galaxies with mean ages of 300-800 Myr and found that the contribution of rapid quenching to the build-up of the red sequence is $\sim 20\%$ at $z\sim1.4$ and $\sim50\%$ at $z\sim2.2$. \cite{Wild2016} also used photometric data (using a super-color selection) to identify post-starbursts and concluded that the post-starburst phase accounts for 25-50\% of the growth of the red sequence at $z\sim1$. In summary, quite a significant fraction of quiescent galaxies can be explained by the PSB phase \citep[e.g.,][]{Snyder2011, Wild2016}, especially at higher redshifts, highlighting the importance of the rapid quenching phase.

In this study we focus on a population of young quiescent galaxies at $z\sim1.5$ (with inferred mean stellar ages below 300 Myr), which has not been explored before, to understand the rapid quenching process at high redshifts and the significance of the rapid quenching phase in galaxy evolution. In Section~\ref{sec:data}, we describe the photometric data we use and how we select 12 rapidly quenched candidates at $z\sim1.5$ based on their rest-frame color. We also describe the Magellan/FIRE spectroscopic observation we conducted on a subset of our sample. In Section~\ref{sec:results}, we present results about their star formation histories, sizes, and information we can learn from emission lines detected from spectroscopic observation.
In Section~\ref{sec:the_role_of_rapid_quenching}, we estimate how many quiescent galaxies have gone through the rapid quenching phase based on the crossing time of the rapid quenching region. 
In Section~\ref{sec:simulation}, using the TNG100 simulation, we identify rapidly quenched analogs and study what caused the starburst and rapid quenching at high redshifts. Finally, in Section~\ref{sec:discussion}, we discuss in more detail the size evolution of young quiescent galaxies and what it suggests, and also the possible physical mechanisms responsible for this rapid quenching, and discuss the overall picture of quenching at high redshifts. The summary and conclusion of our work is given in Section~\ref{sec:summary_conclusion}.

\section{Data and sample selection} \label{sec:data}
\subsection{Rapidly quenched candidates selection from the UltraVISTA catalog}
At high redshifts ($z>1$), it is very challenging to identify spectroscopically-confirmed PSBs, as it requires much more time to detect the Balmer absorption lines that indicate the presence of young stellar populations. Therefore, we follow the method used in \cite{Belli2019} where they inferred the mean stellar ages based on the location in the $UVJ$ diagram (i.e., rest-frame $U-V$ versus $V-J$ colors). We select the youngest quiescent galaxies with inferred ages $< 300\,\rm Myr$, which are the most likely to be rapidly quenched. We describe below in more detail how we select the parent sample and our rapidly quenched candidates.

\begin{figure*}
    \centering
    \includegraphics[width=0.65\textwidth]{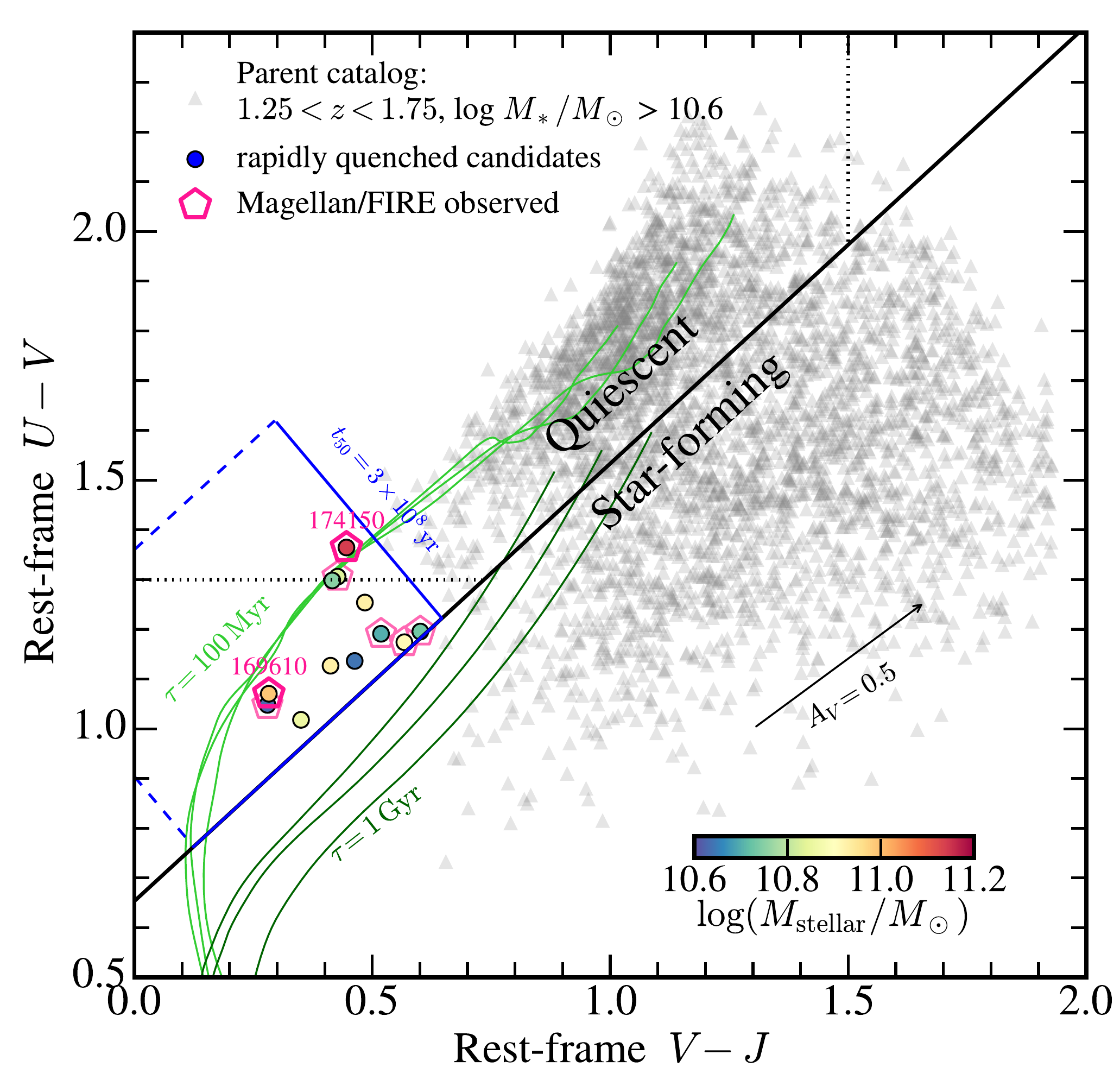}
    \caption{Rest-frame $UVJ$ color-color diagram for selecting rapidly quenched candidates. The $UVJ$ colors are from the UVISTA catalog, where the rest-frame colors are calculated with the EAZY code \citep{Brammer2008}. The gray triangles are the parent sample of massive galaxies ($\log(M_{\rm stellar}/M_\odot)>10.6$ at $1.25<z<1.75$) from the UVISTA catalog. The diagonal black line divides galaxies into quiescent and star-forming galaxies, and the dashed lines are additional constraints used in \cite{Muzzin2013}. The black arrow shows the effect of dust attenuation in the $UVJ$ space, assuming the \citet{Calzetti2000} extinction law. We apply the method used in \cite{Belli2019} and identify 12 rapidly quenched candidates (youngest non-star-forming galaxies), which are marked as filled circles and are color-coded by their stellar mass. We perform Magellan/FIRE observations of seven of these rapidly quenched candidates, marked as magenta pentagons. We highlight two galaxies, UVISTA 169610 and 174150, with their IDs, for which we have detected absorption features. The lime green and green lines are the evolutionary tracks for dust-free stellar population models with star formation history exponentially declining with the $e$-folding timescale of 100 Myr and 1 Gyr, respectively. For each timescale, the three tracks correspond to three different values of stellar metallicity ($\log(Z/Z_\odot)=-0.2, 0.0, 0.2$).}
    \label{fig:UVJ_diagram}
\end{figure*}

We use the COSMOS/UltraVISTA catalog \citep{Muzzin2013a_ApJS} (UVISTA) and select galaxies in the photometric redshift $1.25<z<1.75$ with $\log(M_{\rm stellar}/M_\odot) \ge 10.6$. The redshift range is chosen so that the most important optical absorption and emission line features can be observed in the near-infrared from the ground. We exclude objects that have $K_s > 23.4$, as they are likely to be point sources, and those having bad fits to their photometry (\texttt{chi2} $> 1.5$). Fig.~\ref{fig:UVJ_diagram} shows the parent sample as gray triangles in the rest frame $UVJ$ diagram. 
The $UVJ$ colors are from the UVISTA catalog, where the rest-frame colors are calculated with the EAZY code \citep{Brammer2008}. Following \cite{Belli2019}, we use the rotated coordinates in the $UVJ$ plane \citep[see also][]{Fang2018} defined as follows, which helps us quantify the age trend along the diagonal direction:
\begin{align*}
S_Q =& \,0.75\,(V-J) + 0.66\,(U-V) \\
C_Q =& -0.66\,(V-J) + 0.75\,(U-V).
\end{align*}
Then, the median stellar age of galaxies ($t_{\rm 50}$) can be inferred from the $UVJ$ colors as follows: $\log(t_{50}/\rm yr) = 7.03 + 1.12 * S_{Q}$. \citet{Belli2019} calibrated this approximate relation using a spectroscopic sample with stellar ages older than 300 Myr. In this work we aim at investigating rapidly quenched candidates defined as the galaxies having $t_{50} < 300 \,\rm Myr$ and $C_Q > 0.49$ (quenched). The blue box in Fig.~\ref{fig:UVJ_diagram} indicates the selection region for rapidly quenched galaxies (the dashed lines being arbitrary cuts).

Out of the 3595 objects in the parent sample, 12 galaxies satisfy our selection criteria. The 12 selected galaxies are shown in Fig.~\ref{fig:UVJ_diagram} as filled circles, color-coded by their stellar mass \citep[taken from][]{Muzzin2013a_ApJS}. The lime green and green lines are the evolutionary tracks for dust-free stellar population models generated with the stellar population synthesis code FSPS \citep{Conroy2009}, assuming an exponentially declining star formation history (SFH) with an $e$-folding timescale of 100 Myr and 1 Gyr, respectively. For each timescale we show three tracks corresponding to $\log(Z/Z_\odot)=-0.2, 0.0, 0.2$. These tracks are generated with dust-free stellar models, and the presence of dust would shift each track towards the red (see black arrow in Fig.~\ref{fig:UVJ_diagram}). This means that, at the location of each track, observed galaxies can be substantially younger and dustier than these simple models suggest. Thus the $e$-folding timescales of 100 Myr and 1 Gyr represent the upper bounds of quenching timescales at each location. Indeed, the selected 12 galaxies are located between these evolutionary tracks, indicating that they are most likely quenched very rapidly.

\subsection{Structural data from 3D-DASH survey}
We use the data from 3D-DASH survey \citep{Mowla2022} to explore the morphology and sizes of our rapidly quenched UVISTA galaxies. The 3D-DASH program is a Hubble Space Telescope WFC3 F160W imaging and G141 grism survey targeting the COSMOS field, with an efficient Drift And SHift (DASH) observing technique \citep{Momcheva2017}. 
Global structural parameters for 3D-DASH, including Sersic indices and sizes, are measured using GALFIT \citep{Peng2002GALFIT} identically to the methods in \cite{Cutler2022}.

\subsection{FIRE observations and data reduction}
We conducted the observations with a long-slit Echelle mode, generally using a 0.6''-wide slit, which corresponds to a spectral resolution of $\sigma=50\,\rm km\,s^{-1}$, with a fixed position angle of 0 deg. For each observed galaxy, we aimed to have an exposure time of $\approx 4\,\rm hrs$, and we used the high-gain mode (1.2e-/DN). To improve the sky subtraction, we performed an A-B dithering mode for integration times of $\approx900$ seconds each.

The \texttt{FIREHOSE} pipeline\footnote{https://github.com/jgagneastro/FireHose$\_$v2} was used for the data reduction, which traces the orders and applies flat fielding, wavelength solution, illumination correction, and slit tilt correction. Some A0V stars close to the targets were observed for the telluric correction, which was applied using the \texttt{xtellcorr} package \citep{Vacca2003} implemented in \texttt{FIREHOSE}. The 2D spectrum is extracted from \texttt{FIREHOSE} for each A/B dithering position.

We were able to observe 7 out of the 12 rapidly quenched galaxies; the observations are summarized in Table.~\ref{table:Magellen}. In four cases we detect a noisy stellar continuum but are unable to identify robust features. The lack of emission lines in these four cases suggests that galaxies are not actively forming stars and are likely quenched. In the other three cases we identify emission lines and/or absorption lines.

\begin{table*}[ht!]
\centering
\begin{tabular}{
p{0.06\textwidth}>{\centering\arraybackslash}
p{0.09\textwidth}>{\centering\arraybackslash}
p{0.05\textwidth}>{\centering\arraybackslash}
p{0.03\textwidth}>{\centering\arraybackslash}
p{0.13\textwidth}>{\centering\arraybackslash}
p{0.07\textwidth}>{\centering\arraybackslash}
p{0.10\textwidth}>{\centering\arraybackslash}
p{0.11\textwidth}>{\centering\arraybackslash}
p{0.11\textwidth}>{\centering\arraybackslash}
p{0.05\textwidth}}
\\
\hline
\hline
ID & $\log(M_{\rm \star}/M_\odot)$ & $H$ mag (AB) & $z_{\rm phot}$ & observed & exposure & seeing & emission & absorption & $z_{\rm spec}$ \\
\hline
77854 & 10.70 & 20.6 & 1.34 & Feb 2020 & 2.8 hrs & $\sim$0.6'' & [NII], [OIII] & - & 1.333
\\
199028 & 10.73 & 21.1 & 1.67 & Feb 2020 & 2.5 hrs & 0.5-0.6'' & - & - & - 
\\
39507 & 10.83 & 20.6 & 1.52 & Feb 2020 & 2.8 hrs & 0.8-1.0'' & - & - & - 
\\
24523 & 10.63 & 20.9 & 1.64 & Jan 2021 & 2.5 hrs & $\sim$0.8'' & - & - & - 
\\
169610 & 10.99 & 20.2 & 1.72 & Jan 2021 & 4.0 hrs & 0.4-0.6'' & [NII] doublet & Balmer series &
1.7015 
\\
174150 & 11.14 & 20.2 & 1.72 & Feb-Mar 2022 & 10.3 hrs{}\textsuperscript{\textdagger} & 0.5-0.8'' & No emission & Balmer series & 1.7335 
\\
95964 & 10.91 & 20.4  & 1.50 & Mar 2022 & 4.0 hrs & 0.5-0.6'' &  - & - & -  
\\
\hline
\end{tabular}
\label{table:Magellen}
\vspace{0.1cm}
\begin{minipage}{\linewidth}
    \footnotesize
    {}\textsuperscript{\textdagger} Observed for 3 half nights. For the first two half nights, we used a 0.6’’-wide slit, and we switched to a 0.75'' one on our last observing night. When combining these 3-night data, we smoothed the data of the first two nights to 0.75’’ slit resolution ($\sigma=62.5\,\rm km\,s^{-1}$) and combine it into the third-night data observed with a 0.75'' slit.
\end{minipage}
\caption{Summary of Magellan/FIRE observations of the seven UVISTA galaxies among the 12 rapidly quenched candidates. From the leftmost column, the table lists the ID from UVISTA catalog, stellar mass ($\log(M_{\rm \star}/M_\odot)$), $H$-band magnitude (in AB), photometric redshift ($z_{\rm phot}$), observed dates, total exposure time, seeing, detected emission features, absorption features, and the spectroscopic redshift ($z_{\rm spec}$). Absorption lines are detected in only two galaxies: UVISTA 169610 and 174150. Those two galaxies are highlighted with their IDs in Fig.~\ref{fig:UVJ_diagram}.}
\end{table*}

\section{Results}
\label{sec:results}
\subsection{Star formation history}
\subsubsection{Prospector results using photometry only}
To explore the stellar population properties of the 12 selected galaxies to see if they had a starburst before rapid quenching (thus, whether they are truly ``post-starburst''), we run \texttt{Prospector} \citep{Johnson2021}, a fully Bayesian stellar population inference code, to fit the photometric data released in the UVISTA catalog spanning from FUV to mid-IR \citep[See][for details about the photometric data of the UltraVISTA survey]{Muzzin2013a_ApJS}.

\texttt{Prospector} adopts the stellar population synthesis model \texttt{FSPS} \citep{Conroy2009} to generate synthetic galactic spectral energy distributions. We used MIST isochrones \citep{Choi2016} and assume a Chabrier initial mass function \citep{Chabrier2003GalacticFunction}. The model consists of 19 free parameters describing the contribution of stars, gas, and dust. The nested sampling package Dynesty \citep{Speagle2020} allows us to efficiently sample from the parameter space based on given priors to estimate the Bayesian posteriors. The stellar population of a galaxy is described by a set of parameters, including redshifts, stellar mass, metallicity, dust parameters, and a non-parametric star formation history \citep[see more details about the setup for nonparametric models in][]{Leja2019HowNonparametric, Leja2019AnOlder3DHST}. Dust attenuation is modelled assuming the two-component, birth-cloud component and diffuse component, following \cite{CharlotFall2000}. 
The choice of the prior is very important as the fitting result is sensitive to it.  
We use a continuity prior for the non-parametric SFH, in which we assume that the ratio of SFR between two adjacent time bins follows a Student’s t-distribution with $\sigma=0.3$ and $\nu=2$ \citep{Leja2019HowNonparametric}. The use of continuity prior favors a smooth variation of SFR between the two adjacent time bins and is thus biased against dramatic changes in SFR, such as rapid quenching or starburst. See \cite{Tacchella2022} and \cite{Suess2022} for more details about how the SFH reconstructed from \texttt{Prospector} fitting would be changed when different priors are used. We used 14 time bins for non-parametric SFH, where the earliest bins are 30 Myr and 100 Myr, beyond which bins are evenly spaced in logarithmic ages. A constant SFR is assumed within each time bin.

\begin{figure*}
    \centering
    \includegraphics[width=\textwidth]{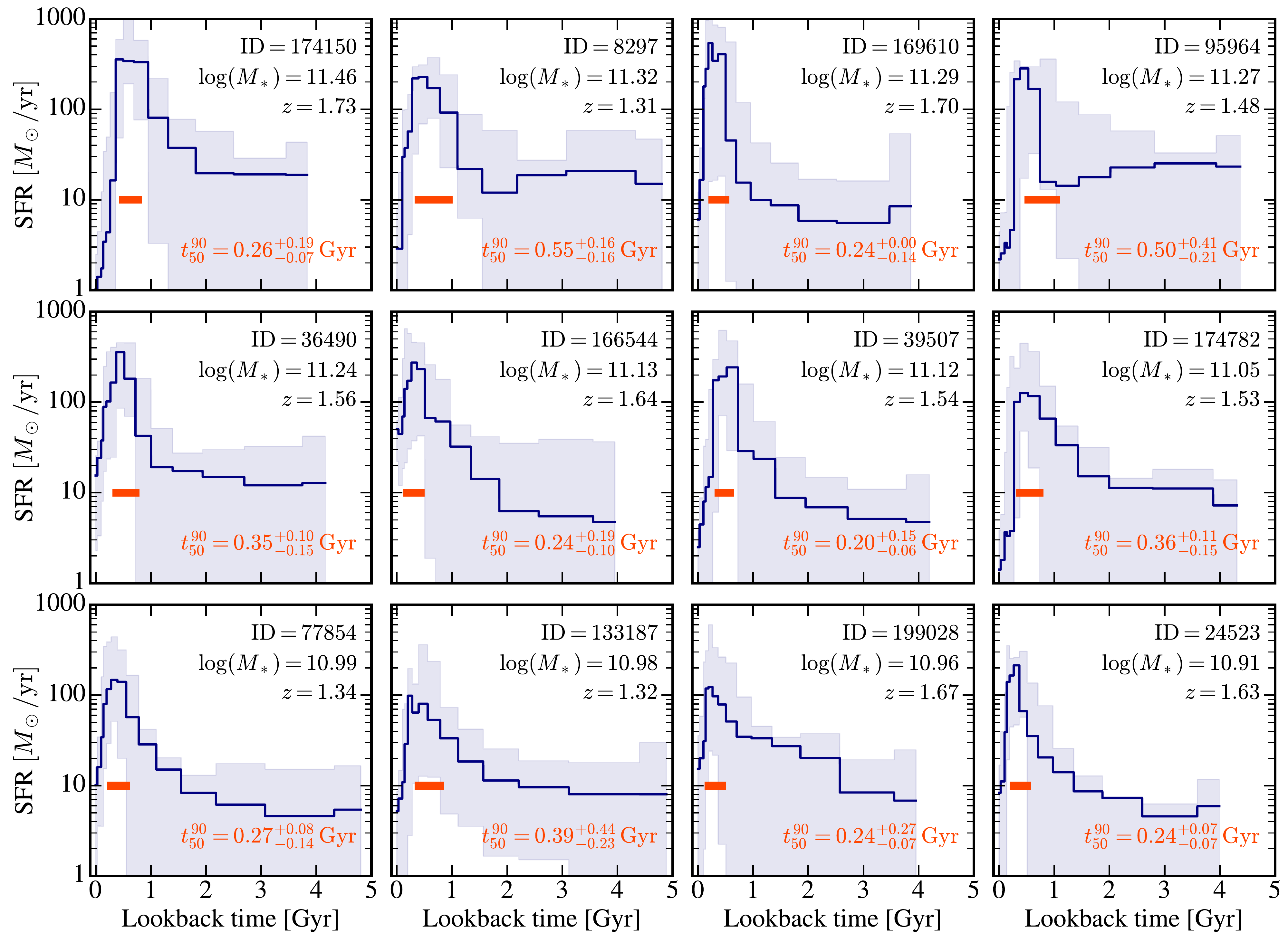}
    \caption{SFH of 12 UVISTA galaxies derived from \texttt{Prospector} fitting on photometric data. The MAP distribution is shown as a solid navy line, while the shade indicates 95\% of the SFH distribution from 1000 random posteriors. The orange horizontal bar indicates the formation timescales ($t_{50}^{90}$, the time it takes for the stellar mass of a galaxy to increase from 50\% to 90\% of the total stellar mass).
    We find that all of our 12 rapidly quenched candidates, selected solely based on their location in the $UVJ$ diagram, are post-starburst galaxies that are rapidly formed (with average formation timescale of $t_{50}^{90}\sim 320\,\rm Myr$).}
    \label{fig:SFH_UVISTA_phot_only}
\end{figure*}

Fig.~\ref{fig:SFH_UVISTA_phot_only} shows the resulting SFH for the 12 objects, reconstructed from \texttt{Prospector} fitting. The solid navy line shows the SFH from the maximum a posteriori probability (MAP), and the shade represents 95\% of the posterior distribution from 1000 random posteriors. 
Indeed, many of our sample galaxies are rapidly quenched with significant starbursts. UVISTA 166544 appears to be not fully quenched with $\rm \log(sSFR)\sim-9.4$.
To quantify how rapidly our sample galaxies are formed, we measure the formation timescale ($t_{50}^{90}$), defined as the time it takes for a galaxy to increase its stellar mass from 50\% to 90\% of the final stellar mass. The orange horizontal bar in each panel indicates the formation timescale of each galaxy. The formation timescale we define here traces the second half of the formation history, which can be better constrained by the observations. Table~\ref{tab:prospector_fit_phot_only} summarizes the \texttt{Prospector} fitting results and the formation timescales $t_{50}^{90}$. The average formation timescale of our 12 targets is $t_{50}^{90}=320\,\rm Myr$, which clearly shows that they are not just rapidly quenched but also rapidly formed. We point out that this is a conservative result because the continuity prior is biased against abrupt changes in the SFHs. The true formation timescales may be even shorter than the values we measure.

\begin{table*}[ht!]
\centering
\begin{tabular}{
p{0.09\textwidth}>{\centering\arraybackslash}
p{0.09\textwidth}>{\centering\arraybackslash}
p{0.07\textwidth}>{\centering\arraybackslash}
p{0.09\textwidth}>{\centering\arraybackslash}
p{0.09\textwidth}>{\centering\arraybackslash}
p{0.09\textwidth}>{\centering\arraybackslash}
p{0.09\textwidth}>{\centering\arraybackslash}
p{0.09\textwidth}>{\centering\arraybackslash}
p{0.11\textwidth}}
\\
\hline
\hline
ID & $\log(M_{\rm \star}/M_\odot)$ & $z_{\rm fitted}$ & $\log(Z/Z_\odot)$ & $\hat{\tau}_{2}^1$ & $n^2$ & $t_{50}^{90}\,\rm \,[Myr]$ & $R_{\rm eff}$ [kpc] & $t_{\rm transition}\,\rm \,[Myr]$ \\
\hline 
174150 & $11.46^{+0.02}_{-0.02}$ & $1.73$ & $0.11^{+0.05}_{-0.09}$ & $0.10^{+0.07}_{-0.04}$ & $-0.32^{+0.31}_{-0.32}$ & $260^{+190}_{-70}$ & $1.06\pm0.03$ & $308^{+41}_{-45}$ 
\\
169610 & $11.29^{+0.03}_{-0.04}$ & $1.70$ & $0.05^{+0.10}_{-0.13}$ & $0.14^{+0.07}_{-0.05}$ & $-0.28^{+0.33}_{-0.19}$ & $240^{+0}_{-140}$ & - &$408^{+67}_{-76}$  
\\
8297 & $11.32^{+0.03}_{-0.05}$ & $1.31$ & $0.00^{+0.11}_{-0.26}$ & $0.05^{+0.03}_{-0.02}$ & $-0.63^{+0.21}_{-0.19}$ & $550^{+160}_{-160}$ & - &$243^{+115}_{-60}$ 
\\
36490 & $11.24^{+0.03}_{-0.02}$ & $1.56$ & $0.10^{+0.06}_{-0.09}$ & $0.29^{+0.05}_{-0.05}$ & $-0.51^{+0.14}_{-0.19}$ & $350^{+100}_{-150}$ & - &$154^{+53}_{-46}$ 
\\
95964 & $11.27^{+0.02}_{-0.05}$ & $1.48$ & $-0.36^{+0.39}_{-0.21}$ & $0.07^{+0.11}_{-0.05}$ & $-0.49^{+0.27}_{-0.32}$ & $500^{+410}_{-210}$ & $1.98\pm0.22$ &$414^{+48}_{-162}$ 
\\
166544 & $11.13^{+0.07}_{-0.05}$ & $1.64$ & $-0.08^{+0.14}_{-0.48}$ & $0.43^{+0.09}_{-0.08}$ & $-0.12^{+0.14}_{-0.15}$ & $240^{+190}_{-100}$ & - &$112^{+57}_{-70}$  
\\
39507 & $11.12^{+0.03}_{-0.03}$ & $1.54$ & $0.04^{+0.09}_{-0.10}$ & $0.19^{+0.06}_{-0.06}$ & $-0.58^{+0.27}_{-0.24}$ & $200^{+150}_{-60}$ & - &$315^{+53}_{-63}$ 
\\
174782 & $11.05^{+0.03}_{-0.03}$ & $1.53$ & $0.08^{+0.08}_{-0.12}$ & $0.21^{+0.08}_{-0.07}$ & $-0.57^{+0.26}_{-0.21}$ & $360^{+110}_{-150}$ & $0.94\pm0.13$ &$235^{+60}_{-53}$ 
\\
199028 & $10.96^{+0.07}_{-0.04}$ & $1.67$ & $0.11^{+0.06}_{-0.18}$ & $0.45^{+0.08}_{-0.10}$ & $-0.44^{+0.09}_{-0.13}$ & $240^{+270}_{-70}$ & - &$47^{+54}_{-26}$ 
\\
77854 & $10.99^{+0.04}_{-0.04}$ & $1.34$ & $0.13^{+0.04}_{-0.09}$ & $0.40^{+0.06}_{-0.06}$ & $-0.36^{+0.12}_{-0.12}$ & $270^{+80}_{-140}$ & $1.14\pm0.16$ &$97^{+42}_{-36}$ 
\\
133187 & $10.98^{+0.03}_{-0.04}$ & $1.32$ & $-0.33^{+0.13}_{-0.18}$ & $0.17^{+0.05}_{-0.05}$ & $-0.61^{+0.18}_{-0.23}$ & $390^{+440}_{-230}$ & $3.05\pm1.74$ &$302^{+98}_{-77}$ 
\\
24523 & $10.91^{+0.03}_{-0.04}$ & $1.63$ & $-0.02^{+0.15}_{-0.19}$ & $0.15^{+0.04}_{-0.04}$ & $-0.71^{+0.14}_{-0.13}$ & $240^{+70}_{-70}$ & $0.86\pm0.04$ &$363^{+97}_{-75}$ 
\\
\hline
\end{tabular}
\begin{minipage}{\linewidth}
    \footnotesize
    {}\textsuperscript{1} The parameter describing the optical depth for the diffuse dust component \citep[see details in][]{Conroy2009} \\
    {}\textsuperscript{2} The power-law modifier to the shape of the \cite{Calzetti2000} dust attenuation curve \citep[see details in][]{Kriek_Charlie2013}.\\
\end{minipage}
\vspace{-0.2cm}
\caption{\texttt{Prospector} fitting results of 12 UVISTA rapidly quenched galaxies based on their photometry, the formation timescale ($t_{50}^{90}$), effective radius ($R_{\rm eff}$), and the rapid quenching transition time ($t_{\rm transition}$, defined in Section~\ref{sec:rapid_quenching_transition_time}) derived from the resulting SFH. The galaxies are in descending order of stellar mass from the \texttt{Prospector} fits.}
\label{tab:prospector_fit_phot_only}
\end{table*}

While fitting models to the photometric data gives us a rough idea of how rapidly galaxies are quenched, the detailed quenching history, as well as whether galaxies are truly quenched or showing any AGN signatures, can only be revealed with spectroscopic data. 

\subsubsection{Prospector results using both photometry and spectroscopy}
We detect clear absorption features for two galaxies, UVISTA 169610 and 174150 – two of the most massive galaxies in our sample. To study their stellar population properties in more detail, we perform \texttt{Prospector} fitting again using both the photometric data from the UVISTA survey and the spectroscopic data that we obtained from the Magellan/FIRE observations. When fitting a spectrum, the velocity dispersion of a galaxy is used as an additional free parameter. Fitting both photometry and spectroscopy requires calibration when combining the two information; we follow the common approach of multiplying a polynomial function with the model spectrum to match the observed spectrum \citep[see the details about the spectrophotometric calibration in][]{Johnson2021}. The order of the polynomial is another additional free parameter and we set it to 10. For spectrum fitting, we mask out emission lines and bad pixels.

\begin{figure*}
\centering
\renewcommand{\arraystretch}{1.8}
\setlength{\tabcolsep}{0pt}
\begin{tabular}{c}
\includegraphics[width=1.0\textwidth]{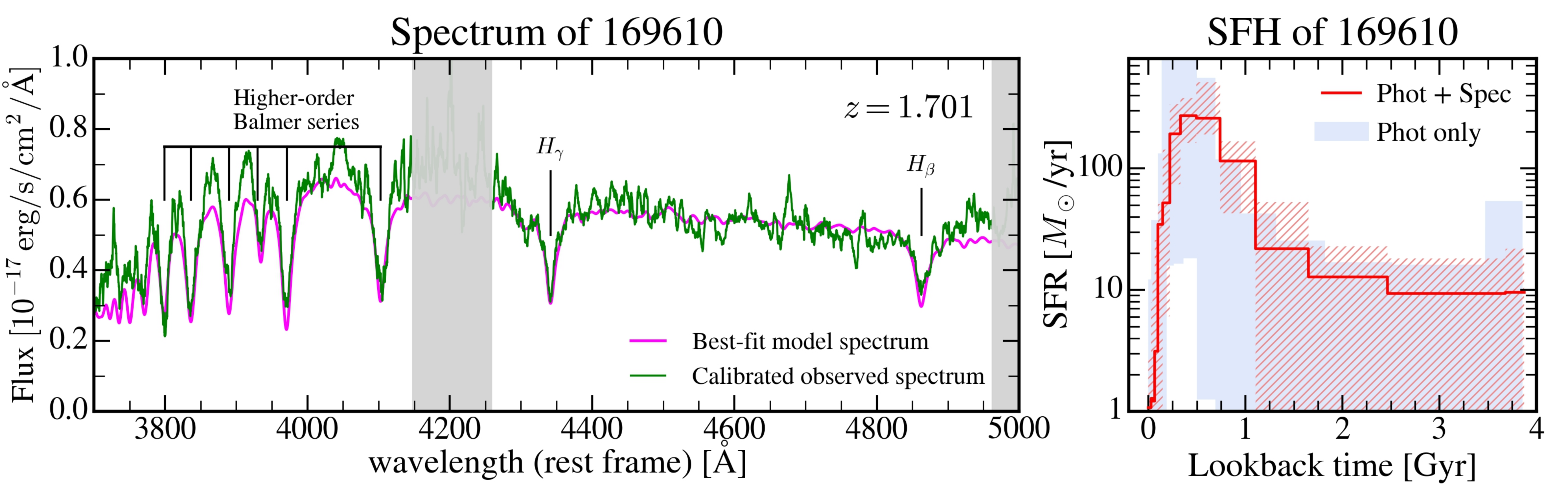} \\
\includegraphics[width=1.0\textwidth]{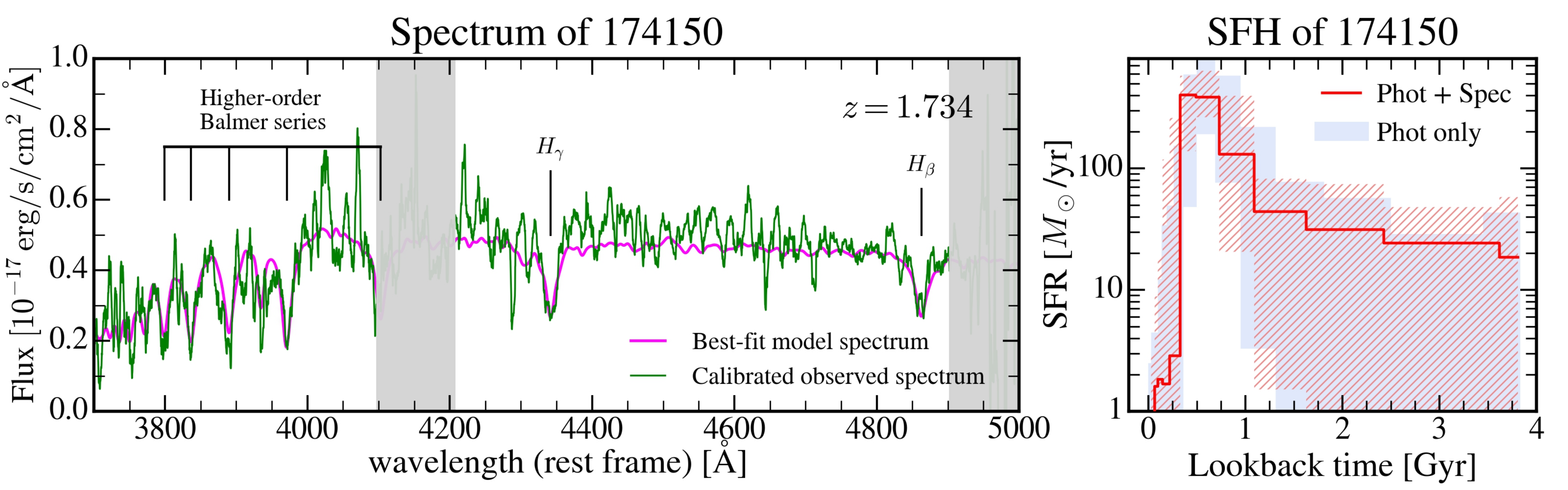} 
\end{tabular}
\caption{Magellan/FIRE spectroscopic data of UVISTA 169610 (top) and UVISTA 174150 (bottom). The smoothed data (smoothed with 29 pixels, corresponding to $\rm \sim15\AA$) is in green in both panels. The magenta lines show the best-fit stellar models from \texttt{Prospector} fitting. The right panels show the resulting SFH of each galaxy. The red line shows the MAP distribution, and the hatched region includes the 95\% of the posterior distribution. The blue shade in the background shows the 95\% of the posterior distribution of SFH from \texttt{Prospector} fitting on photometric data only (same as shown in Fig.~\ref{fig:SFH_UVISTA_phot_only}). We confirm that both galaxies are rapidly formed with starbursts $\sim500\,\rm Myr$ before being rapidly quenched.}
\label{fig:FIRE_data_and_SFH}
\end{figure*}

Fig.~\ref{fig:FIRE_data_and_SFH} shows the Magellan/FIRE spectra of UVISTA 169610 (top) and UVISTA 174150 (bottom). The magenta lines on the left panels show the best-fit models from \texttt{Prospector} fitting, which match both the spectra and the photometry, in physical units of $\rm erg\,s^{-1}\,cm^{-2}\,\AA^{-1}$. The green lines are the observed spectra smoothed by a Gaussian with a width of 29 pixels ($\rm \sim15\AA$); in order to account for the necessarily imperfect flux calibration, the spectra are multiplied by a slowly-varying polynomial so that their continuum matches the best fit. The gray windows are the wavelength regions dominated by skylines and/or large errors. Each panel on the right shows the resulting SFH. The MAP distributions are shown as red lines with hatched regions indicating 95\% of the posterior distributions. As a comparison, we include the posterior distribution when fitting the photometry only (same as Fig.~\ref{fig:SFH_UVISTA_phot_only}) as blue shaded regions.

The new fits yield SFHs that are consistent with those obtained without the spectra.
The error bars of the SFH for UVISTA 169610 have significantly reduced when the spectroscopic data is included for fitting. On the other hand, in the case of UVISTA 174150, the SFH does not seem to be significantly improved with spectroscopic data.
Overall, for rapidly quenched galaxies where the Balmer break is strong, fitting only photometric data appears to be quite effective in constraining the SFH.

\subsection{Sizes}
\begin{figure*}
    \centering
    \includegraphics[width=\textwidth]{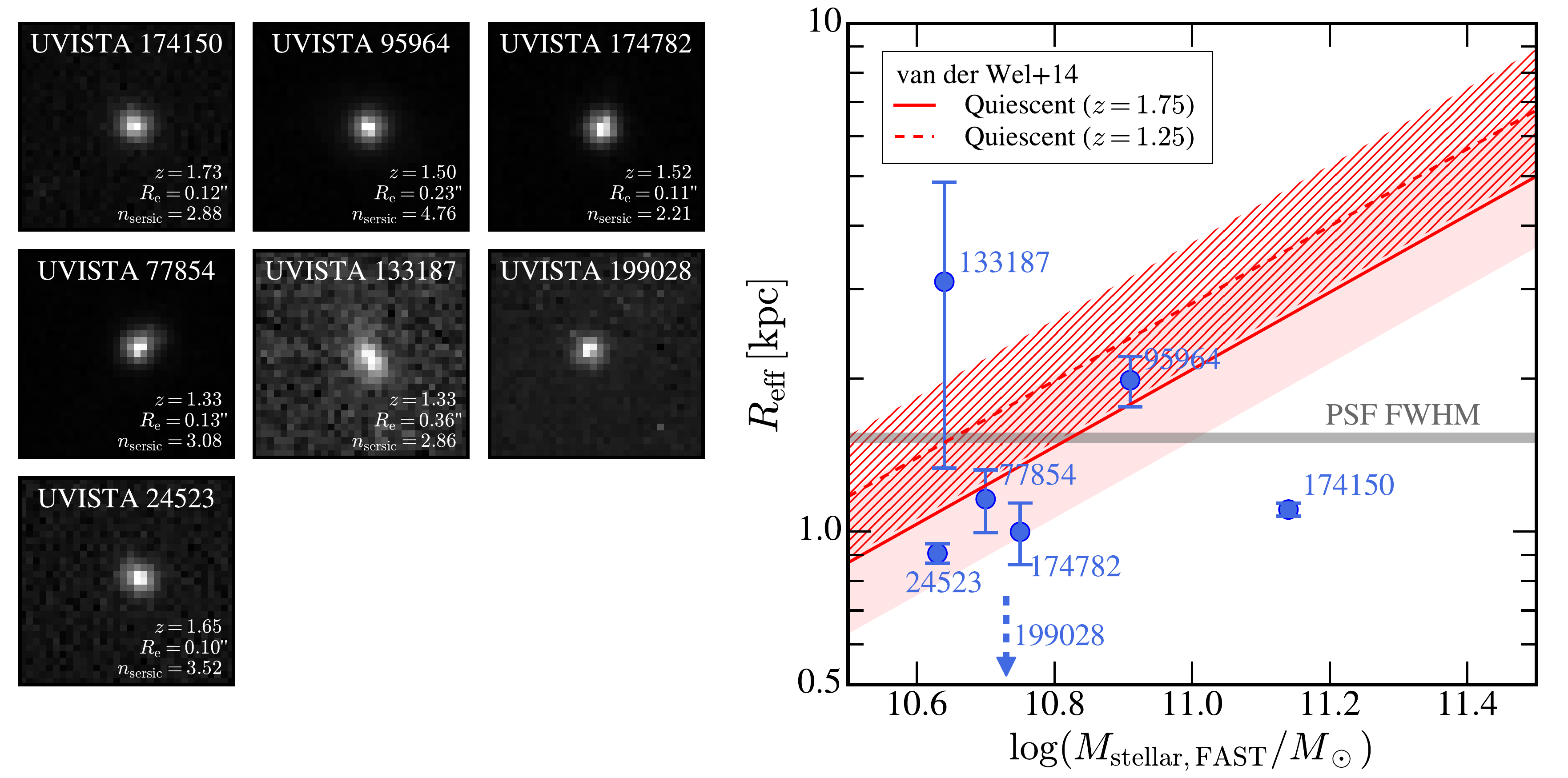}
    \caption{(Left) 3D-DASH cutout images of 7 of our 12 rapidly quenched candidates that are covered in the 3D-DASH survey. The box size is 3 arcsec. The redshift of the galaxy ($z$), its effective radius ($R_{\rm e}$), and the S{\'{e}}rsic index ($n_{\rm Sersic}$) are shown in lower right corner of each panel. UVISTA 199028 appears to be extremely compact, resulting in a bad fit using GALFIT. (Right) Size-mass relation for 6 of our UVISTA rapidly quenched candidates. UVISTA 199028 is plotted as a dashed arrow. The gray horizontal line indicates the PSF ${\rm FWHM}=0.18''$, below which the size of the galaxy might not be well resolved (the measured size being an upper limit). The red dashed line and the hatch region indicate the size-mass relation for quiescent galaxies at $z=1.75$ from \cite{VanDerWel20143D-HST+CANDELS:3}, and the relation for quiescent galaxies at $z=1.25$ is shown as the red solid line and the shade. The stellar mass of galaxies in this plot is derived using \texttt{FAST}, provided in \cite{Muzzin2013a_ApJS}, and sizes are corrected to ``rest-frame'' sizes, to be consistent with \cite{VanDerWel20143D-HST+CANDELS:3}. We find that most of our UVISTA rapidly quenched candidates with the size measurements seem to be more compact than normal quiescent galaxies at their respective redshifts.}
    \label{fig:size_mass}
\end{figure*}

The left panels of Fig.~\ref{fig:size_mass} shows the post stamp images for the 7 out of 12 UVISTA rapidly quenched candidates that are covered by the 3D-DASH survey. The redshift and the GALFIT fitting results for each galaxy, including the effective radius $R_{\rm e}$ and the S{\'{e}}rsic index $n_{\rm Sersic}$, are shown in the lower right corner of each panel. UVISTA 199028 is covered in the survey but appears to be extremely compact, resulting in a bad fit using GALFIT. 

The size-mass relation for the 6 of our 12 UVISTA rapidly quenched candidates are shown in the right panel of Fig.~\ref{fig:size_mass}. UVISTA 199028, which has a bad GALFIT fit, is plotted as a dashed arrow. The red dashed line and the hatch region indicate the size-mass relation for (rest-frame $UVJ$ color-selected) quiescent galaxies at $z=1.75$ from \cite{VanDerWel20143D-HST+CANDELS:3}, and the relation for quiescent galaxies at $z=1.25$ is shown as the red solid line and the shade. The relations presented in \cite{VanDerWel20143D-HST+CANDELS:3} are fits to K-corrected ``rest-frame'' sizes. Thus, for consistency, we also correct the 3D-DASH sizes to ``rest-frame'' sizes, in the same way they did, using the equation (2) in \cite{VanDerWel20143D-HST+CANDELS:3}. Also, in this plot we use the stellar mass of the UVISTA galaxies provided by \cite{Muzzin2013a_ApJS} which used the package \texttt{FAST} \citep{Kriek2009b}, as in \cite{VanDerWel20143D-HST+CANDELS:3}. Broadly, the stellar masses of our 12 rapidly quenched UVISTA galaxies are estimated to be $\sim 0.3$ dex more massive when fitted by \texttt{Prospector} \citep[see][for a discussion of this effect]{Leja2019AnOlder3DHST}. The gray horizontal line indicates the point-spread function (PSF) ${\rm FWHM}=0.18''$ (converted into kpc at $z\sim1.25-1.75$), below which the size of the galaxy might not be well resolved (the measured size being an upper limit).

Most of the seven UVISTA rapidly quenched galaxies are more compact than normal quiescent galaxies at their respective redshifts, consistent with the results of several previous works where they found that young quiescent galaxies (or PSBs) tend to be more compact than old (or normal) quiescent galaxies \citep[e.g.][]{Whitaker2012compact_PSB, Belli2015, Almaini2017, Maltby2018, Wu2018}. There are two exceptions, UVISTA 95964 and 133187. The size of UVISTA 95964 is consistent with the relation of typical quiescent galaxies of the same redshift. UVISTA 133187 appears to have an extended structure, as shown in the left panel. A possible explanation for the extended structure is that this galaxy may be a rotating disk or have unresolved merging companions, but additional kinematic data is needed to confirm this. As a result of being compact, the S{\'{e}}rsic indices for these 6 galaxies are all quite high ($n_{\rm Sersic}>2.5$). Their compact sizes, even more compact than normal quiescent galaxies, provide another piece of evidence that they had central starbursts before quenching, as we have confirmed from the \texttt{Prospector} SED fitting. We will discuss in more detail the possible physical mechanisms behind rapid quenching and compaction in Section~\ref{sec:discussion}.

\subsection{Emission lines}
\label{sec:emission_lines}
AGN activity is thought to play an important role in quenching massive galaxies, but it is very challenging to directly observe it. One of the most common ways to detect AGN activity is through emission-line diagnostics. Many previous studies have shown that local PSBs mostly feature LINER-like emission lines \citep[e.g.,][]{Yan2006, Wild2010, French2015, Alatalo2016b} – although recently it has been pointed out that the low ionization is not necessarily centralized, suggesting that it might be caused by post-AGB stars instead of AGN activity \citep[e.g.,][]{Yan_Blanton_2012, Belfiore2016}. 
At high redshifts, several studies have detected a broad emission component of H$\alpha$ and [NII] in many star-forming galaxies \citep[e.g.,][]{Genzel2014, Forster-Schreiber2019}, which is thought to originate from the inner few kpc. With a very broad kinematics of FWHM$\sim 1000\,\rm km\,s^{-1}$, likely gravitationally unbound, these broad emission components are often associated with ejective AGN feedback.

\begin{figure*}
\centering
\renewcommand{\arraystretch}{0}
\setlength{\tabcolsep}{0pt}

\begin{tabular}{cc}
\includegraphics[width=1.05\columnwidth]{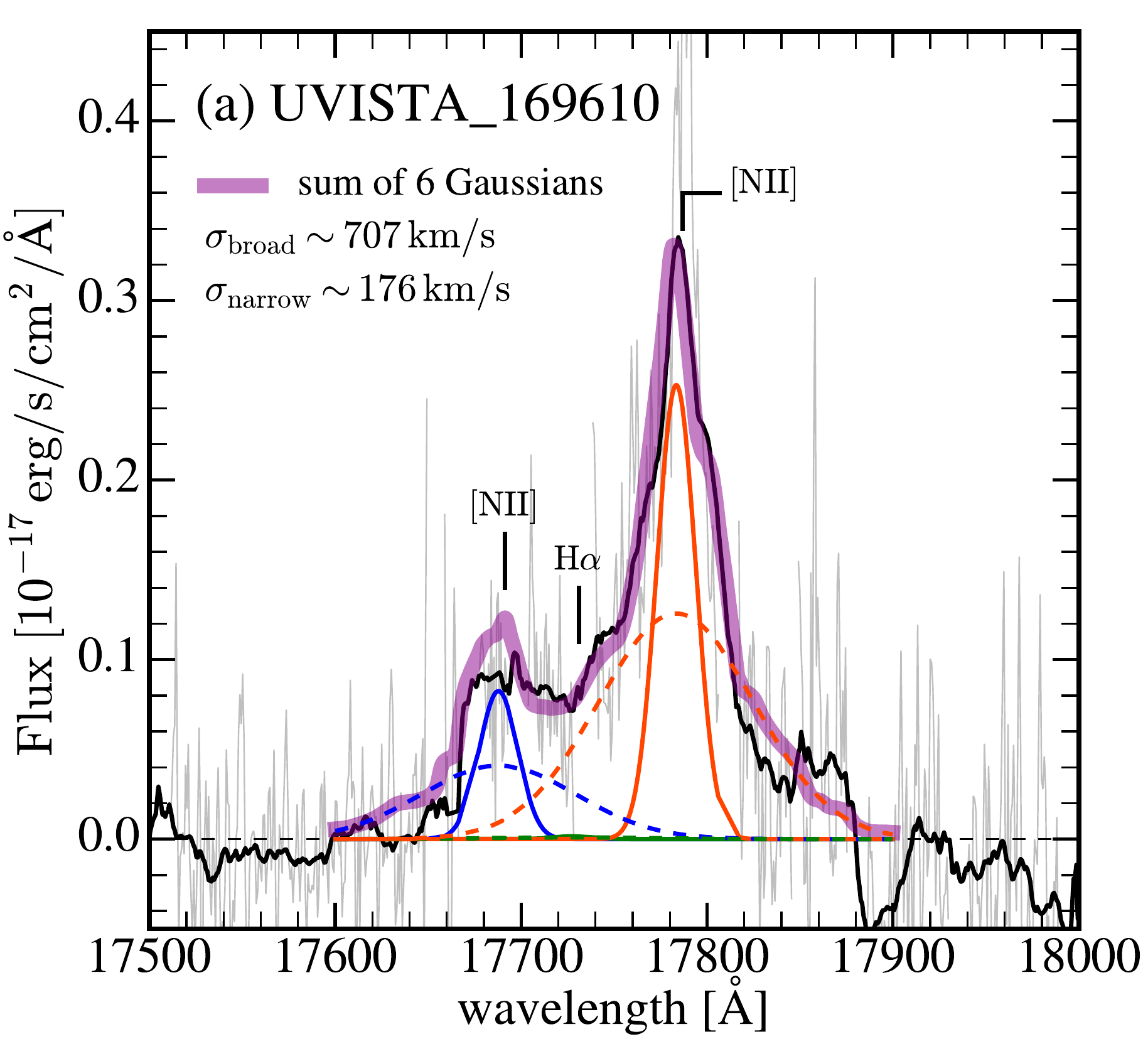} &
\includegraphics[width=1.05\columnwidth]{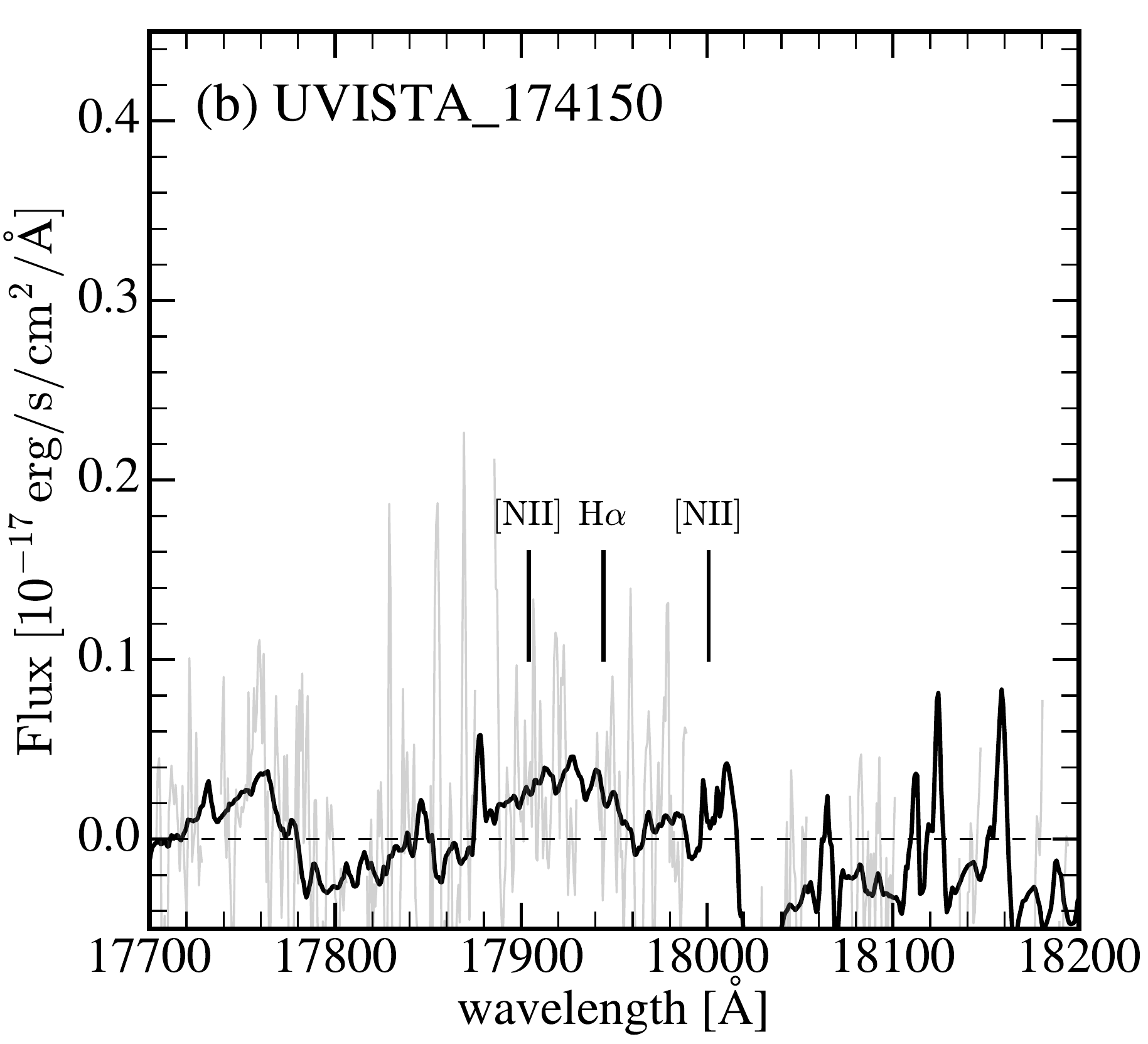} \\
\includegraphics[width=1.05\columnwidth]{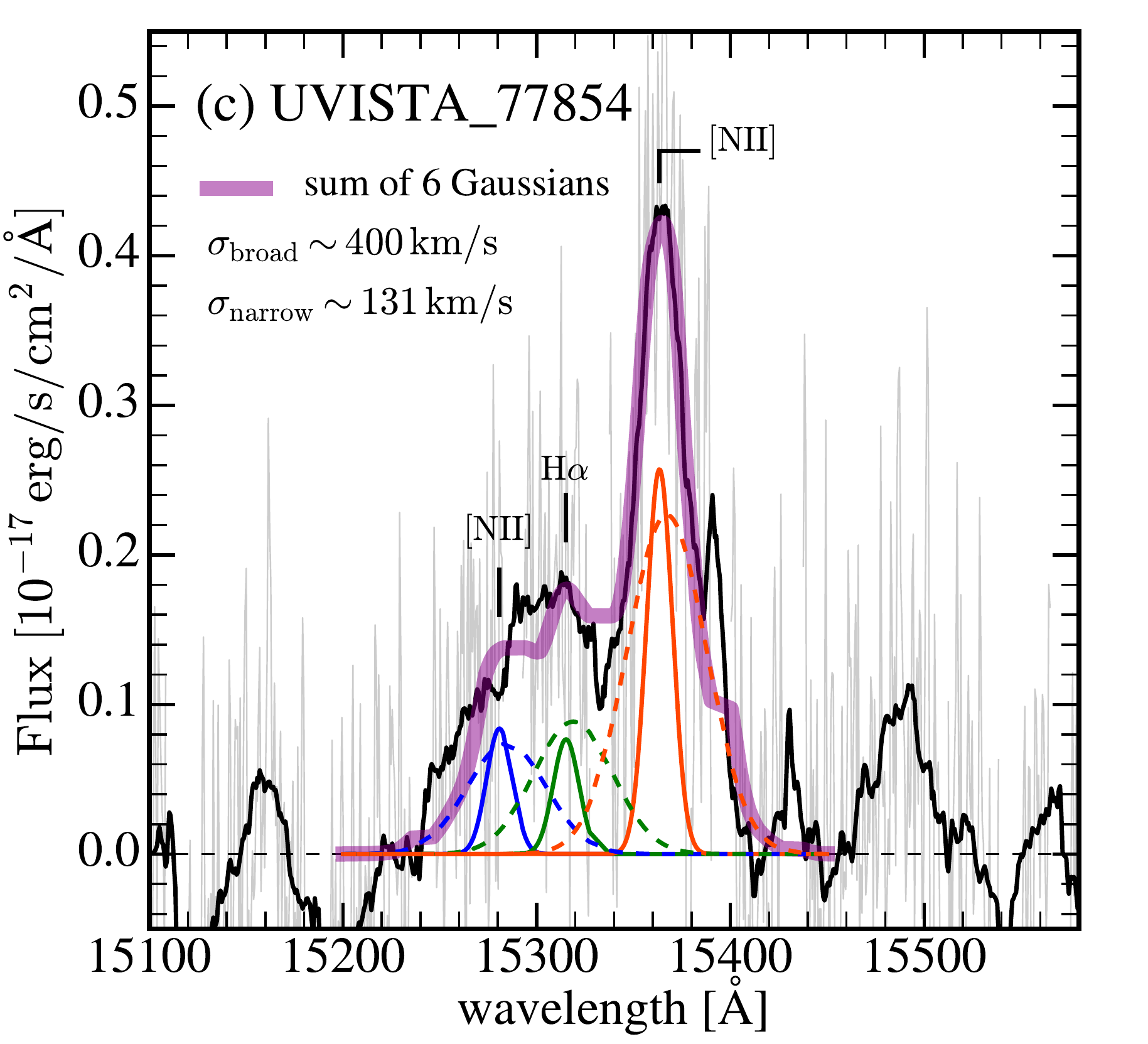} &
\includegraphics[width=1.05\columnwidth]{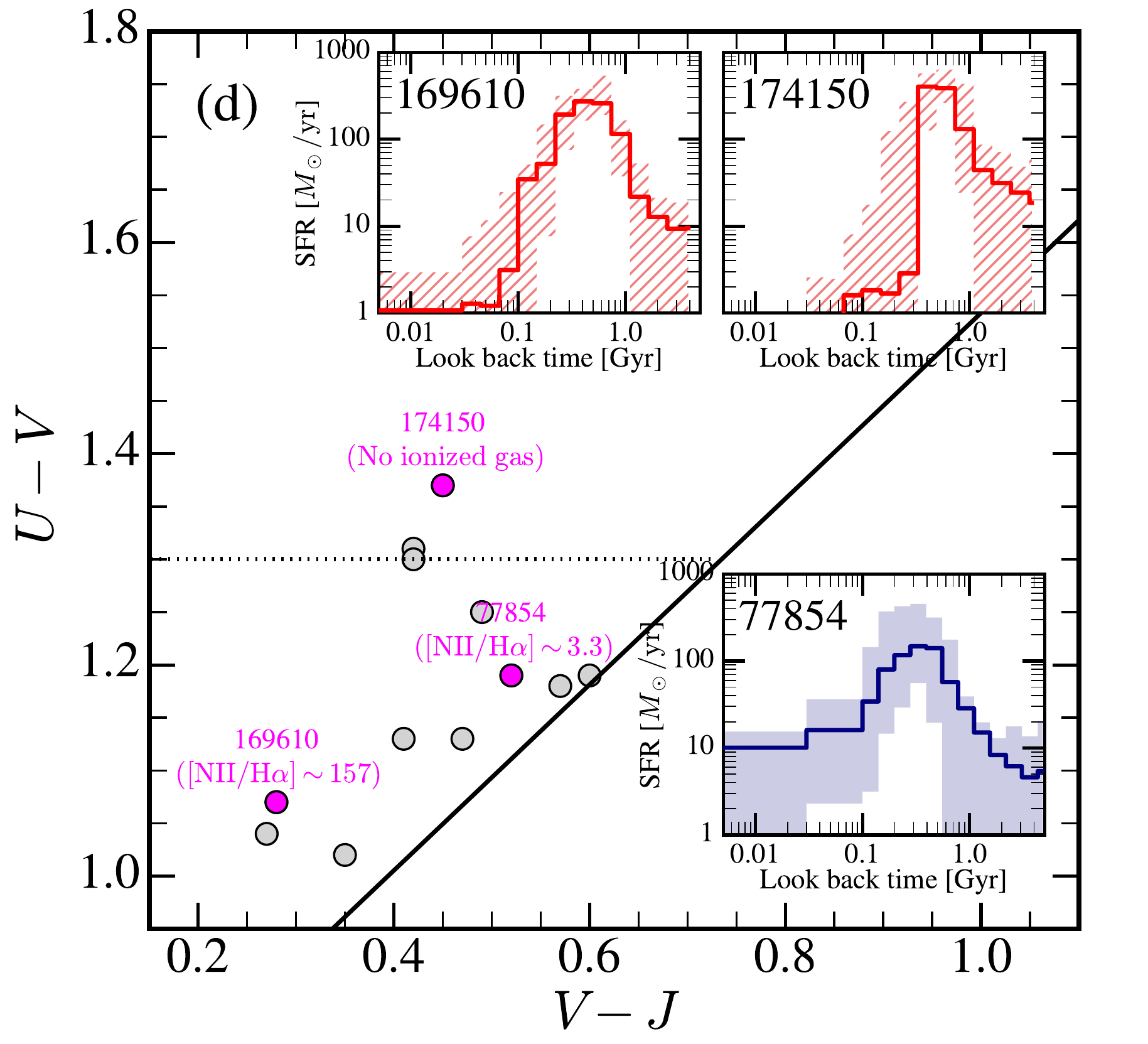}
\end{tabular}

\caption{Ionized gas spectra of (a) UVISTA 169610, (b) UVISTA 174150, and (c) UVISTA 77854 around of H$\alpha$, [NII]$\lambda$6550,$\lambda$6585 emission, in observed wavelength. The raw spectrum is shown in gray lines, and the black solid line in each panel are the ionized gas spectrum smoothed with 29 pixels. For UVISTA 169610 and 77854, we fit 6 Gaussians to the ionized gas spectrum accounting for both broad (solid) and narrow (dashed) line components of H$\alpha$ (green), [NII]$\lambda$6550 (blue), and [NII]$\lambda$6585 (orange). The thick purple line is the sum of the 6 Gaussians, smoothed in the same way as the observed spectrum. For UVISTA 169610, the fitted standard deviations of the broad and narrow line components are $\sigma_{\rm br}=707\,\rm km\,s^{-1}$ ($\rm FWHM_{br}= 2.35\sigma_{\rm br}= 1661\,\rm km\,s^{-1}$) and $\sigma_{\rm nr}=176 \,\rm km\,s^{-1}$, respectively. For UVISTA 77854, the fitted standard deviations are $\sigma_{\rm br}=400\,\rm km\,s^{-1}$ and $\sigma_{\rm nr}=131 \,\rm km\,s^{-1}$. (d) The $UVJ$ locations of the three galaxies and their SFH from \texttt{Prospector} fitting in logarithmic time axis. We find that the ionized outflows found in UVISTA 169610 and 77854 are most likely driven by AGN given their high [NII]/H$\alpha$ ratios and the kinematics of the broad emission components ($\rm FWHM>1000\,\rm km\,s^{-1}$). UVISTA 174150, which is the reddest and quenched earlier, does not show any emission features, which might be because the AGN has already blown away all the ionized gas.}
\label{fig:emission_lines}
\end{figure*}

We obtain the ionized gas spectra of UVISTA 169610 and 174150 by subtracting their best-fit stellar spectrum models from their calibrated observed spectra. Fig.~\ref{fig:emission_lines} (a) and (b) shows the resulting emission line complex of H$\alpha$ and [NII]$\lambda$6550,$\lambda$6585 of UVISTA 16910 and 174150.  For UVISTA 169610, we fit 6 Gaussians to the ionized gas spectrum accounting for both broad and narrow line components of H$\alpha$, [NII]$\lambda$6550, and [NII]$\lambda$6585. The three narrow lines share the same width, and so do all the three broad lines. The flux ratio between ${\rm [NII]}\lambda6550/\lambda6585$ is fixed to be 0.326, following \cite{Forster-Schreiber2019}. We fitted for 7 parameters: the width of broad and narrow line components ($\sigma_{\rm br}$, $\sigma_{\rm nr}$), the line shift between the broad and narrow component centroids ($\Delta\lambda_{\rm shift}$), and the flux of H$\alpha$, [NII]$\lambda$6585 for both broad and narrow line components (H$\alpha_{\rm br}$, [NII]$\lambda6550_{\rm br}$, H$\alpha_{\rm nr}$, [NII]$\lambda6550_{\rm nr}$).

It is clear that there is no ongoing star formation in UVISTA 169610, as indicated by the very low level of H$\alpha$ emission; the measured H$\alpha$ flux is negligible compared to the [NII] flux. From the Gaussian fit of H$\alpha$ emission, we can set an upper limit to the SFR of UVISTA 169610 as SFR $<0.07\,M_\odot/\rm yr$, following the canonical correlation of \cite{Kennicutt1998}, and a lower limit on the ratio of [NII]/H$\alpha> 157.1$. The standard deviation of the broad and narrow line components are $\sigma_{\rm br}=707\,\rm km\,s^{-1}$ ($\rm FWHM_{br}= 2.35\sigma_{\rm br}= 1661\,\rm km\,s^{-1}$) and $\sigma_{\rm nr}=176 \,\rm km\,s^{-1}$, respectively. The kinematics of the broad emission component suggests that the ionized gas outflow is gravitationally unbound, and thus most likely originating from AGN-driven outflows.

The derived ionized gas kinematics is very sensitive to the continuum fitting. For example, when we calibrate the observed spectrum to match the best-fit stellar spectrum using a wavelength window that excludes absorption and emission lines, the fitted kinematics of the broad and narrow line components are $\sigma_{\rm br}= 1200\,\rm km\,s^{-1}$ and $\sigma_{\rm nr}= 191 \rm km\,s^{-1}$. However, this does not change our conclusion that the broad line component is most likely caused by AGN-driven outflows because star formation in this galaxy is so suppressed (as indicated by extremely low H$\alpha$ emission) that it is very unlikely that the broad line component is associated with star formation outflows.

Interestingly, UVISTA 174150 does not seem to show any emission features, neither in H$\alpha$ nor in [NII]. The lack of H$\alpha$ emission can be explained by no ongoing star formation, and the absence of [NII] emission suggests that ionized gas may have already been blown away by the feedback processes. 

We have also detected [NII] and [OIII] emission lines for UVISTA 77854 (no Balmer absorptions detected due to high noise) and confirmed that the spectroscopic redshift is consistent with the phometric redshift. We perform the SED fitting on its photometric data using \texttt{Prospector} with fixed redshift ($z_{\rm spec}=1.333$) and assuming a fixed velocity dispersion of $200\,\rm km\,s^{-1}$. The ionized gas spectrum is obtained by subtracting the best-fit stellar model from the spectrum, which is shown in Fig.~\ref{fig:emission_lines} (c). We also fit the 6 Gaussians, in the same way as we did on UVISTA 169610. The resulting standard deviation of the broad and narrow line components are $\sigma_{\rm br}=400\,\rm km\,s^{-1}$ ($\rm FWHM_{br}= 2.35\sigma_{\rm br}= 940\,\rm km\,s^{-1}$) and $\sigma_{\rm nr}=131 \,\rm km\,s^{-1}$, respectively. The low level of H$\alpha$ emission (setting the upper limit of SFR $<1.13\,M_\odot/\rm yr$ and [NII]/H$\alpha=3.3$) suggests that the galaxy is indeed quenched and the ionized gas outflow is most likely originated from AGN activity.

An interesting trend between the ionized gas content and the $UVJ$ location can be found using these three galaxies (UVISTA 169610, 174150, and 77854), though more galaxies are needed to confirm the trend. Fig.~\ref{fig:emission_lines} (d) highlights their $UVJ$ locations with [NII]/H$\alpha$ ratios. Their SFHs are also shown in the inset panels with the logarithmic time axis. For UVISTA 169610 and 174150, the SFHs from the \texttt{Prospector} fits on both photometry and spectroscopy are shown, whereas, the SFH of UVISTA 77854 is based on photometry fitting (All SFHs are in the same format: MAP shown in solid lines, and hatched or shade area indicating 95\% of posteriors). UVISTA 169610, where H$\alpha$ emisison is almost negligible, resulting in extremely high [NII]/H$\alpha=$157, is indeed in the bluest corner of the $UVJ$ diagram. UVISTA 77854 shows non-negligible level of $H\alpha$ emission, setting the upper limit of SFR $<1\,M_\odot/\rm yr$, which is consistent with the estimated SFR from the \texttt{Prospector} fit. This galaxy is, indeed, closest to the star-forming region in the $UVJ$ diagram. UVISTA 174150 is the reddest galaxy among our 12 rapidly quenched targets, and since it has low amounts of dust, its color is not likely reddened by dust (see the fitting parameter in Table.~\ref{tab:prospector_fit_phot_only}). The SFH of this galaxy also suggests that it was quenched a few hundred Myrs earlier than the other two galaxies, which might be related to the absence of ionized gas. While the other two galaxies still exhibit AGN-driven emission characteristics $\sim100\,\rm Myr$ after quenching, the ionized gas of UVISTA 174150 may have been all blown away $\sim 300$ Myr after quenching.

\section{The role of the rapid quenching phase in the build-up of the quiescent population}
\label{sec:the_role_of_rapid_quenching}
The rapidly quenched galaxies, located at the bluest end of the quiescent sequence on the $UVJ$ diagram, appear to be a rare population even at $z\sim 1.5$. However, they are experiencing a very rapid transition from star-forming to quiescence, and potentially hold important clues about the quenching process. Here, we measure the duration of this rapid quenching phase and, based on this transition timescale, estimate how many quiescent galaxies have gone through the rapid quenching phase.

\subsection{Rapid quenching transition time}
\label{sec:rapid_quenching_transition_time}
To measure how quickly galaxies go through the rapid quenching phase, we explore the evolution of the rest-frame $UVJ$ colors using the SFH we obtained from \texttt{Prospector} fitting. For each galaxy, we extract 1000 random posteriors from the \texttt{Prospector} fitting result and, for each posterior, we generate a stellar population using FSPS with the SFH, metallicity ($\log(Z/Z_\odot)$), $\hat{\tau}_{2}$, and $n$ (dust index) values of the posterior in question. We then calculate the rest-frame $UVJ$ colors of the stellar population generated for every time step. The time steps go beyond the observation epoch, to see how the colors of the rapidly quenched galaxies would evolve in the future (assuming that galaxies remain fully quiescent).

\begin{figure*}
    \centering
    \includegraphics[width=\textwidth]{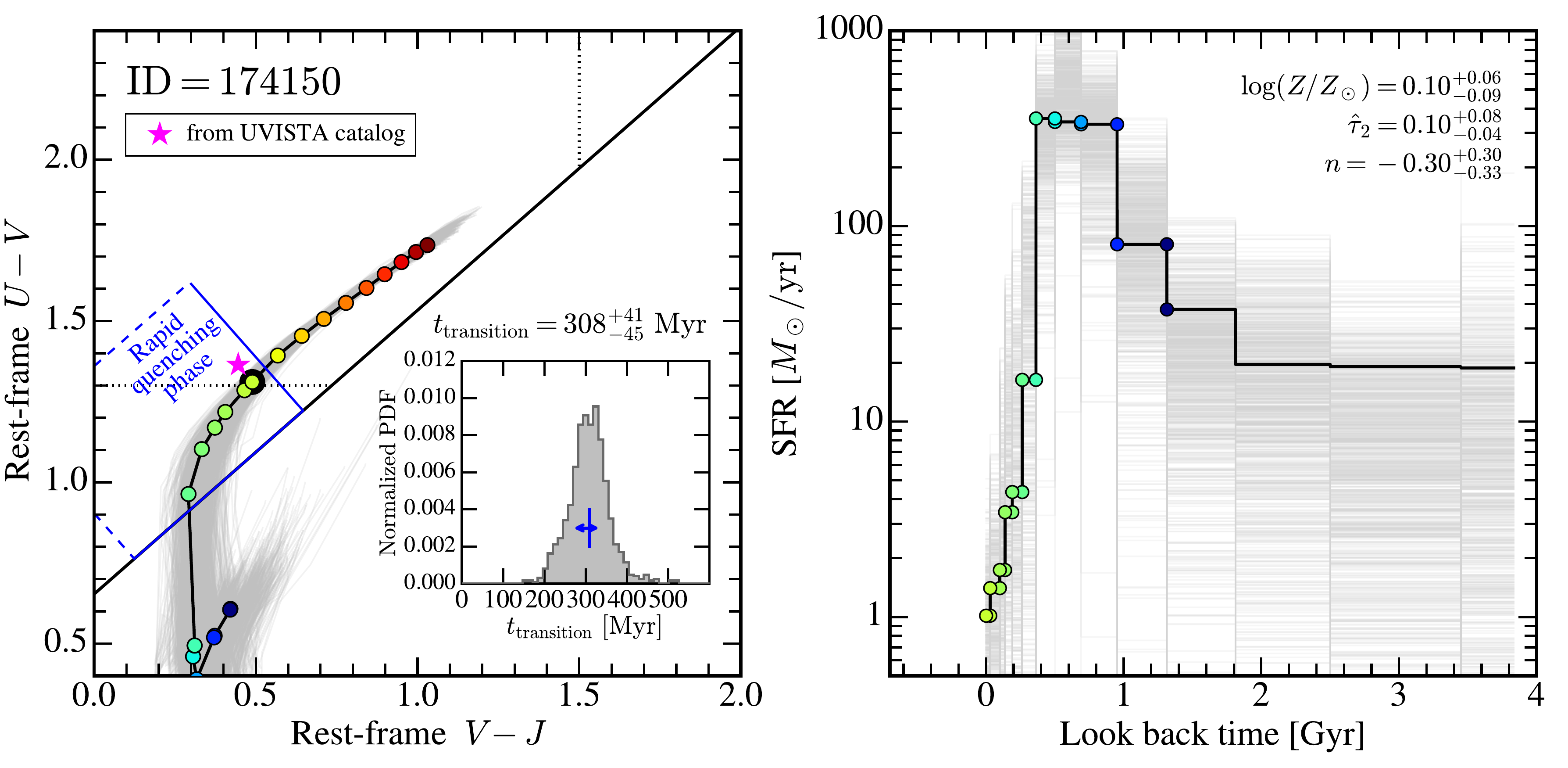}
    \caption{(Left) Rest-frame $UVJ$ color evolution of UVISTA 174150. (Right) SFH from \texttt{Prospector} fitting. In both panels, the results of 1000 random posteriors are shown as gray lines, and the black solid line shows the MAP for visual purposes only (not used for further calculations). Each colored point in both panels indicates a time step, including time steps that go beyond the observed epoch (redder points showing the future color evolution for the next 1 Gyr). The point with a thicker edge in the left panel shows the calculated $UVJ$ colors at the observed epoch (lookback time = 0), and the magenta star shows the $UVJ$ colors provided by the UVISTA catalog. From the \texttt{Prospector} fitting results (SFH and other stellar population parameters), we measure the time it takes for a galaxy to cross the rapid quenching region and define it as the rapid quenching transition time. For UVISTA 174150, the gray histogram in the inset panel on the left shows the distribution of transition time measured with 1000 random posteriors, and the median time for this galaxy is $t_{\rm transition} = 308^{+41}_{-45}\,\rm Myr$.}
    \label{fig:color_evolution}
\end{figure*}

Fig.~\ref{fig:color_evolution} shows the color evolution of one UVISTA galaxy in the $UVJ$ diagram (left) and the SFH from \texttt{Prospector} fitting which we used to calculate the colors (right). For visual purposes, we show the maximum a posterior (MAP) SFH in a solid black line, and the $UVJ$ color evolution based on it. Each colored point in both panels indicates a time step. Each gray line shows the SFH of each posterior and the resulting color evolution. The point with a thicker edge in the left panel shows the calculated $UVJ$ colors at the observed epoch (lookback time = 0), and the magenta star shows the $UVJ$ colors provided by the UVISTA catalog; these are calculated with the EAZY code \citep{Brammer2008}, and may be slightly different from the ones obtained with \texttt{Prospector} (thick black edge).

The blue box in the left panel represents the region which we define as the ``rapid quenching phase'' ($C_q> 0.49$ and $t_{50}<3\times10^8\,\rm yr$). For each of the 12 galaxies in our sample we measure the time it takes to cross that region and define it as the (rapid quenching) transition time ($t_{\rm transition}$). The inset panel shows the normalized PDF of $t_{\rm transition}$ calculated with 1000 individual posterior samples with each SFH and metallicity, $\hat{\tau}_{2}$, $n$ (dust index) fixed during the evolution. The median $t_{\rm transition}$ of 1000 posterior samples for this object is $t_{\rm transition} = 308^{+41}_{-45}\,\rm Myr$, and is shown as the blue vertical line in the inset panel, with blue arrows indicating the 16th to 84th percentile range.

\begin{figure*}
    \centering
    \includegraphics[width=\textwidth]{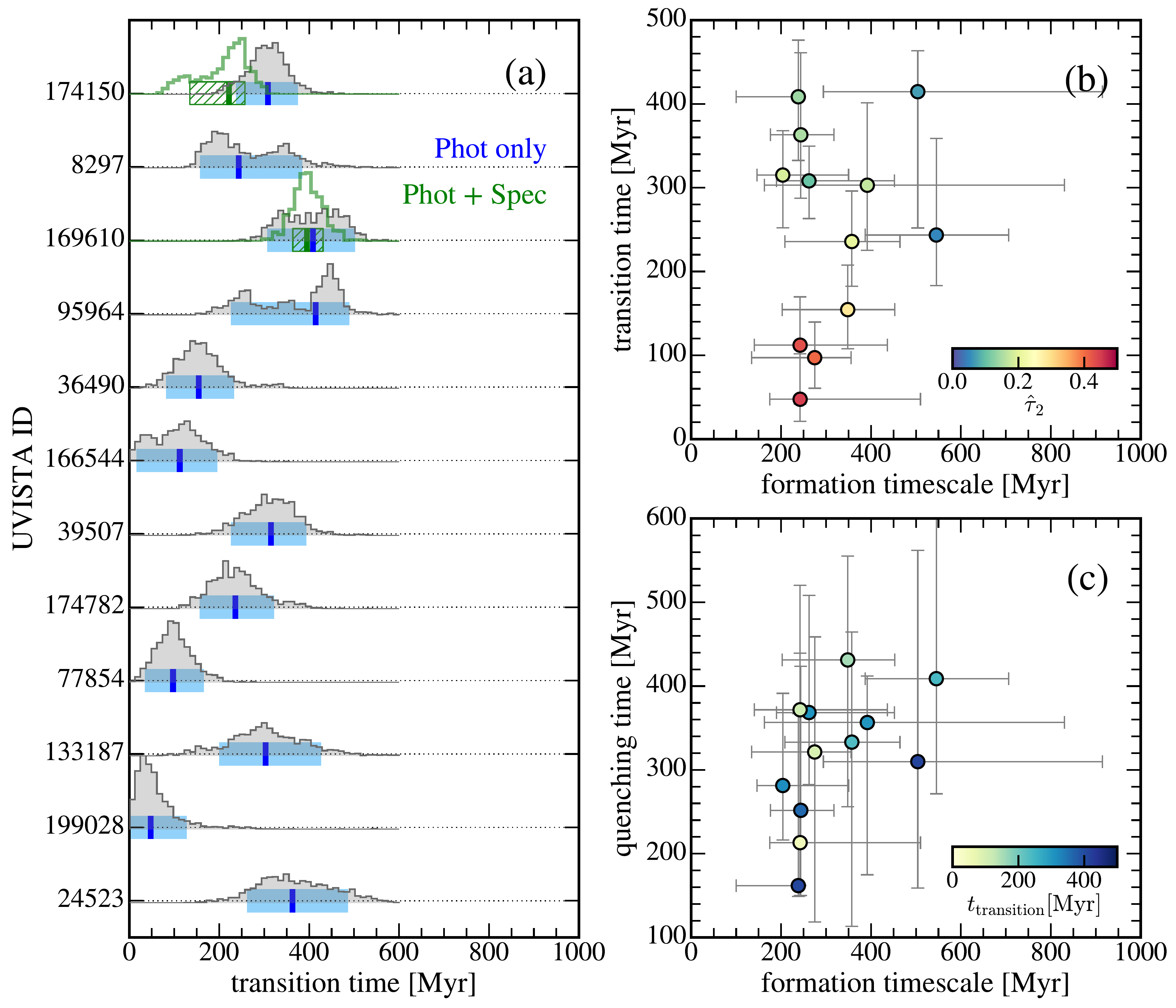}
    \caption{(a) Rapid quenching transition time of 12 UVISTA galaxies. The gray histogram shows the normalized PDF of the transition time of 1000 random posterior samples. The median timescale of 1000 random posterior samples is shown in blue vertical lines, with a sky blue bar indicating the range from 16th to 84th percentile. (b) Comparison between transition time and the formation timescale ($t_{50}^{90}$). (c) Comparison between the quenching timescale and the formation timescale. We find that the average rapid quenching transition time of our samples is $t_{\rm transition} \approx 300 \,\rm Myr$ and the transition time does not appear to be correlated with the formation or quenching timescales.}
    \label{fig:transition_time}
\end{figure*}

Fig.~\ref{fig:transition_time} shows the measured transition time of all 12 UVISTA rapidly quenched galaxies. Each gray histogram shows the normalized PDF of the transition time of 1000 random posterior samples for each galaxy. The median timescale of 1000 random draws is shown in blue vertical lines, with a sky blue bar indicating the range from 16th to 84th percentile. In the case of UVISTA 199028 and UVISTA 166544, with some of their random posteriors, they end up not going through the rapid quenching phase region (the blue box in the left panel of Fig.~\ref{fig:color_evolution}), but rather evolve by deflecting to the right corner of the region. This explains why the lower tail of their transition time distributions seems to go below 0. For these two galaxies, therefore, we measure the median transition timescale only using the random posteriors with which they go through the rapid quenching region. The average (rapid quenching) transition time of the 12 rapidly quenched UVISTA galaxies is $t_{\rm transition} =250 \,\rm Myr$. Excluding the three dust-rich galaxies, the average transition time is $t_{\rm transition} = 305 \,\rm Myr$. This estimate of the average transition time is likely an upper bound, given that we used fixed dust parameters during the evolution and galaxies were almost certainly dustier before quenching. The true transition time may be even shorter than the values we measure.

For UVISTA 169610 and 174150, we measure the transition times using the SFH from the \texttt{Prospector} fitting on both photometric and spectroscopic data (i.e., the SFH presented in Fig.~\ref{fig:FIRE_data_and_SFH}). The green histogram shows the distribution of transition times measured from 1000 random posterior samples, and the median and 16th to 84th percentiles are shown as green vertical lines and hatched bars. For UVISTA 169610, the median transition time is similar to that measured with the fit on photometric data only, yet the error bar is much better constrained with the spectroscopic data. On the other hand, the transition time of UVISTA 174150 has lowered when measured with the fit on both photometry and spectroscopy.

We find that dusty galaxies tend to have short transition timescales, as they cross only the right corner of the rapid quenching region (and in some random posteriors, they do not even cross the region, as in UVISTA 199028 and 166544). This implies that very dusty rapidly quenched galaxies might have not been captured in our selection box, and our estimated rapid quenching transition time holds true for relatively less dusty galaxies. The existence of dusty rapidly quenched galaxies also calls into question the implicit assumption that the amount of dust is a function of time, as they are assumed to be even dustier in the recent past (before quenching).

Fig.~\ref{fig:transition_time} (b) compares the transition time with the formation timescale $t_{50}^{90}$. The formation timescales, plotted as the orange horizontal line in Fig.~\ref{fig:SFH_UVISTA_phot_only}, indicate how rapidly galaxies are formed (time including both starburst and rapid quenching phase). The transition time and formation timescales do not appear to be related as they trace different processes, and also transition times are mostly affected by the amount of dust, as indicated by the color code.

Fig.~\ref{fig:transition_time} (c) compares the formation timescale with the quenching timescale. The quenching timescale here indicates how quickly star formation subsides since starburst, defined as $t_{\rm SF\,peak} - t_{\rm quench}$ where $t_{\rm SF\,peak}$ is epoch when star formation is at its peak and $t_{\rm quench}$ is the quenching epoch. The quenching epoch $t_{\rm quench}$ is defined as the time when specific SFR (sSFR) drops below sSFR = $1/[2\,t_{\rm univ}(z)]$ where $t_{\rm univ}(z)$ is the age of the Universe \citep[e.g.,][]{Tacchella2022_Fast_slow_quenching, Park2022}. There seems to be a weak trend between formation timescale and quenching timescale; galaxies that are formed more rapidly with strong starbursts appear to be quenched more rapidly. The quenching timescale, however, does not correlate with the transition time shown as color code.

In summary, the average (rapid quenching) transition time of our rapidly quenched UVISTA galaxies is $t_{\rm transition} \approx 300 \rm \,Myr$ (excluding the three dust-rich galaxies with very short transition times). This estimate of transition is likely an upper bound, as we used fixed dust parameters during the evolution and galaxies were almost certainly dustier in the past. While the spectroscopic data seems to help constrain the transition time, based on the 2 galaxies, photometric data seems to be sufficient to estimate the transition time. More spectra are needed to determine exactly how much spectroscopic data will change the picture. The transition time seems to be related to the amount of dust (dustier galaxies tend to have shorter transition times), but not with how rapidly galaxies are formed (formation timescales).

\subsection{Number density}
\label{sec:number_density}
\begin{figure*}
    \centering
    \includegraphics[width=\textwidth]{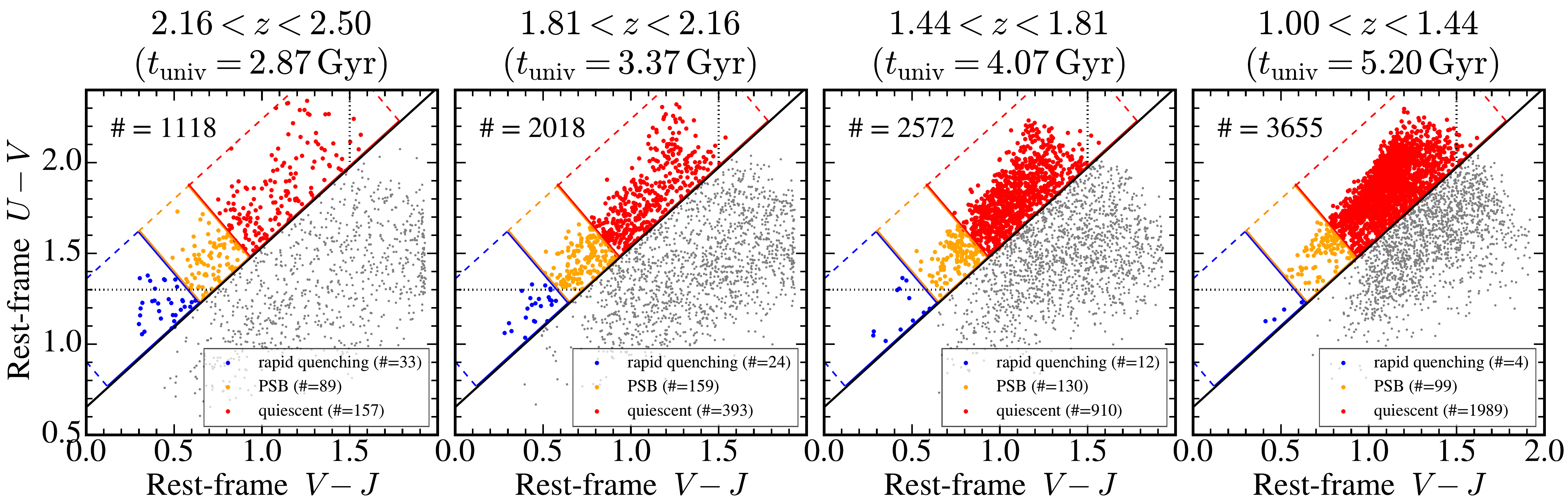}
    \caption{$UVJ$ color-based identification of quiescent (red), PSB (orange), and rapidly quenched (blue) galaxies in four redshift bins of equal comoving volumes. The redshift range and the mean age of the Universe are displayed on top of each panel. The total number of galaxies in each redshift bin is shown at the upper left corner of each panel.}
    \label{fig:number_density}
\end{figure*}

Based on the average rapid quenching transition time of $t_{\rm transition} \approx 300 \rm \,Myr$, we estimate how many quiescent galaxies have gone through this rapid quenching phase.
In the same way as we selected the rapidly quenched galaxies ($C_Q > 0.49$ and $t_{50}<3\times10^8\,\rm yr$), we use the mean stellar age ($t_{50}$) inferred in \cite{Belli2019} to further classify quenched galaxies (above the line of $C_Q > 0.49$) into quiescent ($t_{50} > 8\times10^8\,\rm yr$) and PSB ($3\times10^8 < t_{50}\rm /yr < 6\times10^8$) galaxies. Fig.~\ref{fig:number_density} shows the resulting quiescent (red), PSB (orange), and rapidly quenched galaxies (blue) in the $UVJ$ diagram in four redshift bins that have the same comoving volume by construction. We apply the same mass cut ($\log(M_{\rm stellar}/M_\odot) > 10.6$) for all redshift bins. The total number of galaxies in each redshift bin is shown in the upper left corner of each panel.

\begin{figure}
    \centering
    \includegraphics[width=\columnwidth]{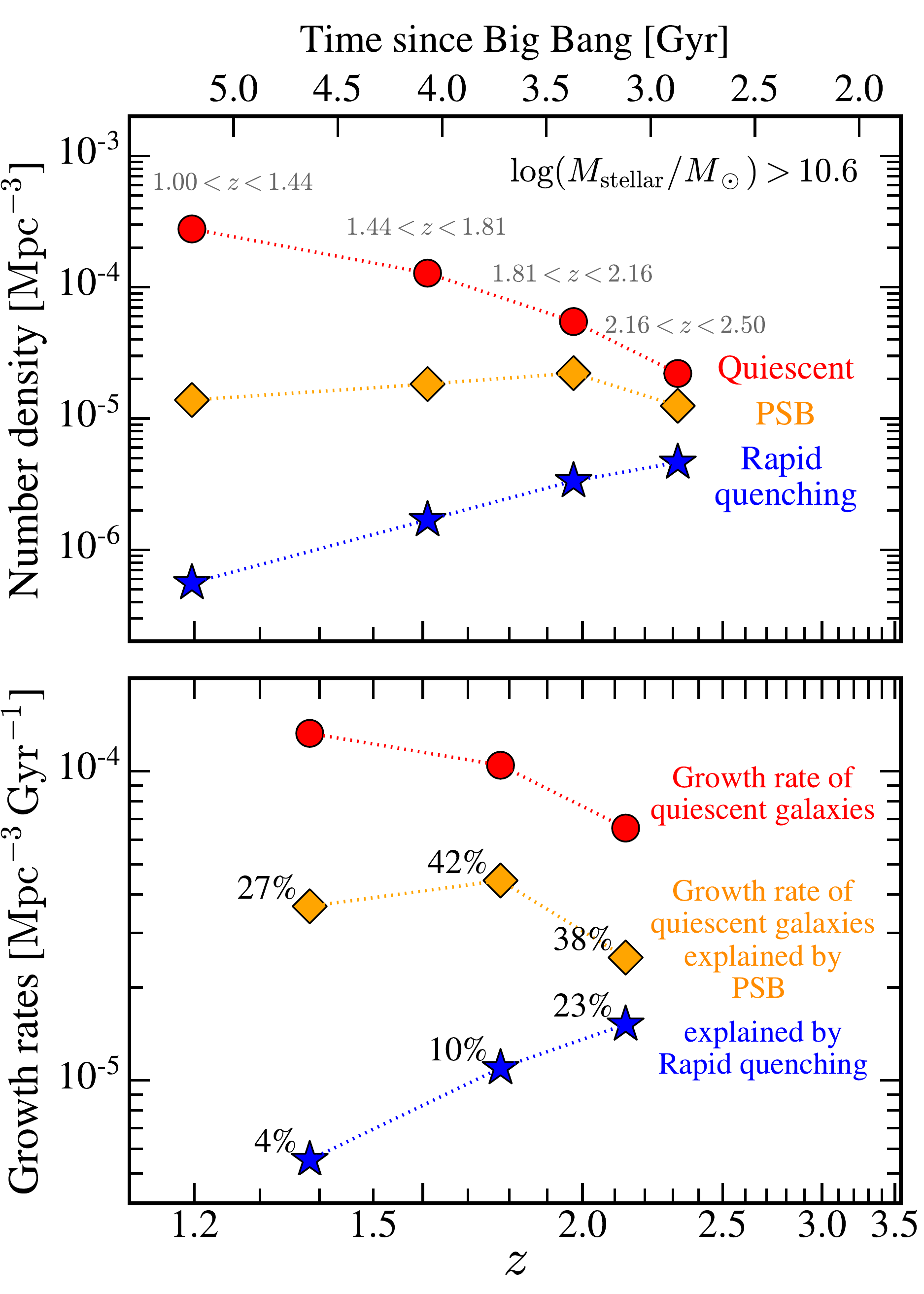}
    \caption{(Top) Number density of quiescent (red), PSB (orange), and rapidly quenched (blue) galaxies as a function of redshift. (Bottom) Growth rates of the quiescent population in red circles. The growth rates of quiescent galaxies which can be explained by the transition from PSB and from rapid quenching phases are plotted in orange diamonds and blue stars (percentages displayed next to the symbols). 
    We find that the fraction of quiescent galaxies that have gone through the rapid quenching phase increases with redshift: 4\% at $z\sim1.4$, 10\% at $z\sim1.8$, and 23\% at $z\sim2.2$.}
    \label{fig:number_density2}
\end{figure}

The top panel of Fig.~\ref{fig:number_density2} shows how the number density of quiescent (red), PSB (orange), and rapidly quenched (blue) galaxies changes with redshift. While the number density of quiescent galaxies increases with time, the number density of PSB and rapidly quenched galaxies (the young quiescent populations) decreases significantly with time. This redshift trend of the number density of old and young quiescent galaxies is consistent with previous work \citep[e.g.,][]{Whitaker2012, Wild2016, Belli2019}.
The growth rate of the quiescent population is calculated by the difference of their number densities in two adjacent redshift bins divided by the time interval between the two bins, and is plotted as red circles in the bottom panel.

We calculate how much of the growth of the quiescent population can be explained by the transition of galaxies through the rapid quenching phase (assuming galaxies remain quiescent after quenching). This can be calculated by dividing the number density of rapidly quenched galaxies at the previous epoch by their transition time, which is $t_{\rm transition}^{\rm rapid\,quenching}= 300\,\rm Myr$. We calculate the contribution of PSB galaxies to the growth of the quiescent population in an analogous way, assuming a transition time $t_{\rm transition}^{\rm PSB}\sim 500\,\rm Myr$ (the time it takes for a passive evolution from $t_{50}=300\,\rm Myr$ to $800\,\rm Myr$), following \cite{Belli2019}. Note that since rapidly quenched galaxies will eventually evolve into the PSB regions, the contribution rates of PSB and rapid quenching calculated here are inclusive.

The growth rate of the quiescent population that can be explained by the transition from the PSB and the rapid quenching phases is plotted as orange diamonds and blue stars, respectively, in the bottom panel of Fig.~\ref{fig:number_density2}. Only 4\% of the growth of quiescent galaxies can be explained by the transition from the rapid quenching phase at $z\sim1.4$, suggesting that not many quiescent galaxies seem to have gone through this rapid quenching phase at this redshift. However, the rapid quenching phase seems to account for higher fractions of quiescent galaxies at higher redshifts ($23\%$ at $z\sim2.2$). This is consistent with the fact that in the early universe galaxies need to be quenched rapidly in order to be identified as quiescent galaxies, given the short amount of available time. Note that the estimated contributions of rapid quenching to the growth of the quiescent population are lower bounds, as the rapid quenching transition time we measured ($t_{\rm transition}=300\,\rm Myr$) is likely an upper bound.

Dry mergers of quiescent galaxies can also contribute to the growth rate of the quiescent population, which we have neglected in our analysis. The impact of dry mergers on the number density of a population can have two opposite effects; if the two (massive) galaxies in the selected quiescent population merge into one galaxy, the number density decreases. Whereas, a dry merger of two slightly less massive galaxies results in one massive quiescent galaxy, which introduces a new galaxy in the population, thus increasing the number density. \citet{Belli2019} have assessed the impact of these two competing effects and concluded that the two effects almost cancel out for galaxies more massive than $\log(M_{\rm stellar}/M_odot)>10.8$. Since we have used a similar mass threshold for our number density calculation, the effect of dry mergers would not affect our conclusion.

\section{Rapid quenching in the TNG100 simulation}
\label{sec:simulation}
To understand what caused the starburst and rapid quenching at high redshifts in more detail, we use the TNG100 simulation to see if rapidly quenched analogs can be reproduced in simulations, and if they exist, to study how they are formed.

The TNG100 simulation, as part of the Illustris TNG project \citep{Springel2018MNRAS475, Naiman2018MNRAS477, Marinacci2018MNRAS480, Pillepich2018FirstGalaxies, Nelson2018FirstBimodality, Nelson2019ComAC}, is a magneto-hydrodynamical cosmological volume simulation run using the {\sc Arepo} code. The box size of the simulation is $\sim 100\,\rm Mpc$ (comoving), and the baryonic and dark matter (DM) mass resolution is $m_{*}= 1.4\times10^6\,M_\odot$ and $m_{\rm DM}=7.5\times10^6\,M_\odot$, respectively. We use \texttt{hydrotools} \citep{Diemer2017, Diemer2018ApJSHydrotools} to extract the data from the simulation.

Here we give a brief description about the feedback prescriptions implemented in the TNG models. Detailed descriptions can be found in \cite{Pillepich2018TNGmodels, Weinberger2017MNRAS465, Weinberger2018, Pillepich2021}. In the TNG models, stellar particles are formed in a gas cell where the density is above the threshold density of $n_{\rm}\simeq0.1\,\rm cm^{-3}$ following the Kennicutt-Schmidt relation \citep{Springel2003MNRAS339}, and the Chabrier initial mass function \citep{Chabrier2003GalacticFunction} is assumed for each stellar particle. As the stellar population evolves with time, it returns mass and metals into the surrounding medium by AGB winds (for stars with masses of $1-8\,M_\odot$), and supernovae Type II ($8-100\,M_\odot$) and Ia. A super-massive black hole (SMBH) with a mass of $\sim10^6\,M_\odot$ is seeded at the potential minimum of a halo more massive than $7.4\times10^{10}\,M_\odot$ and it grows either via SMBH-SMBH mergers or by accretion following the Bondi-Hoyle accretion rate. The AGN feedback is modelled in two ways, depending on the accretion rate relative to the Eddington rate, and only one of the two modes is turned on at a time. When the accretion rate is high, thermal energy is isotropically and continuously injected into the surrounding medium (thermal or quasar mode), whereas when the accretion rate is low, feedback energy is injected in the form of kinetic energy in a pulsed and directed way (kinetic or wind mode).

\begin{figure*}
    \centering
    \includegraphics[width=\textwidth]{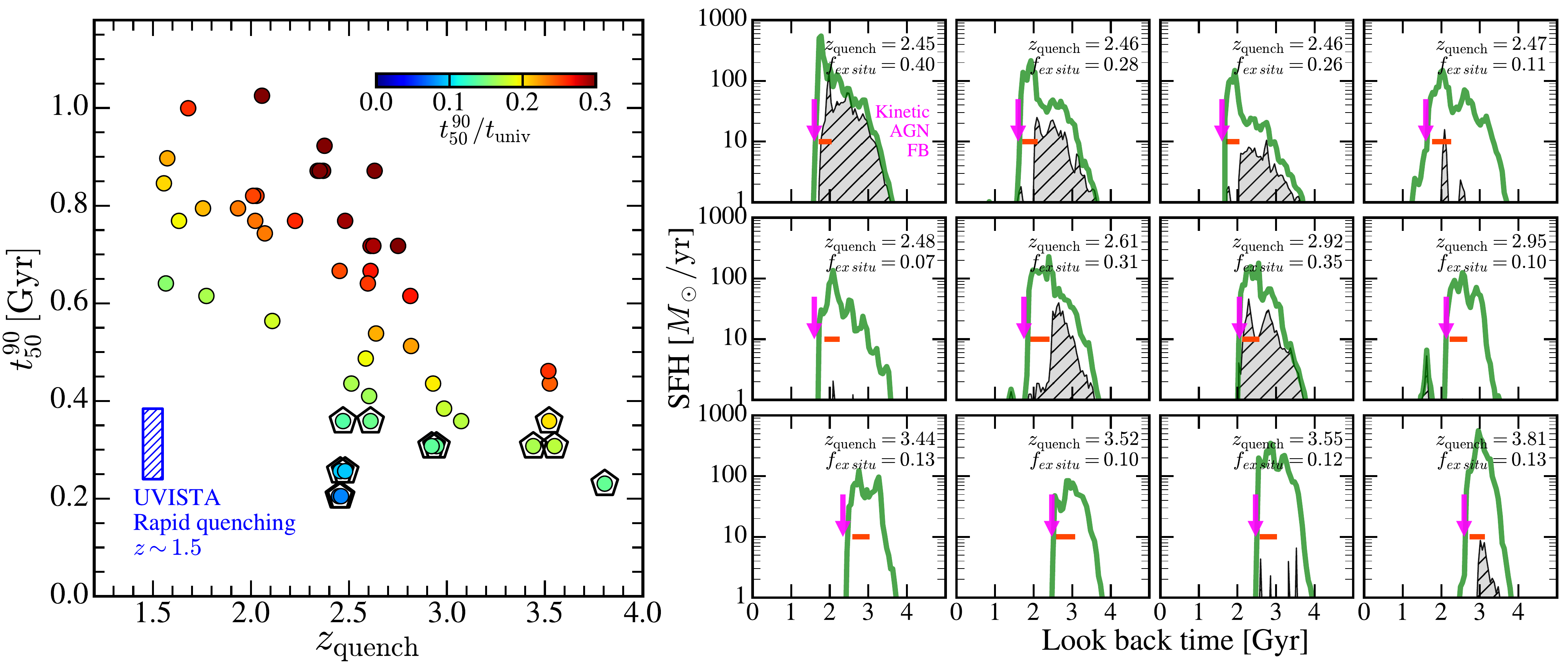}
    \caption{(Left) Formation timescale $t_{50}^{90}$ of massive quiescent TNG100 galaxies ($10.6<\log(M_{\rm stellar}/M_\odot)<12.0$) at $z\sim1.5$ as a function of their quenching epoch ($z_{\rm quench}$). Galaxies are color-coded by their $t_{50}^{90}$ normalized by the age of the Universe at their quenching epoch. For comparison, the blue hatched bar indicates the 16th-84th percentile range of the $t_{50}^{90}$ of 12 rapidly quenched UVISTA galaxies. We select 12 quiescent TNG100 galaxies that have shortest $t_{50}^{90}$, marked in pentagons. 
    (Right) SFH of 12 TNG100 galaxies at $z\sim1.5$ that are most rapidly formed. The green solid line shows the SFH derived from the age distribution of all stellar particles within $3\,R_{\rm eff}$, and the hatched grey histogram shows the SFH of stellar particles formed ex situ and later accreted. The formation timescales are shown as orange horizontal bars. The quenching epoch $z_{\rm quench}$ and the mass fraction of the stellar particles formed ex situ $f_{ex\,situ}$ are given in the right corner of each panel. We mark the epoch when the kinetic AGN feedback is turned on for each galaxy in magenta arrow. It seems clear that kinetic AGN feedback is the key driver of the rapid quenching in TNG100. While major mergers appear to have triggered the starbursts in 5 of 12 rapidly quenched analogs in TNG100, the starbursts in the rest are likely triggered by other mechanisms not involving major mergers. 
    }
    \label{fig:sfh_tng100_rapid_formation}
\end{figure*}

There are in total 918 central galaxies with stellar mass $10.6<\log(M_{\rm stellar}/M_\odot) < 12.0$ at $z=1.5$ in TNG100. Since dust is not included in the simulation, instead of $UVJ$ colors, we select non-star-forming (or passive) galaxies by applying the cut of sSFR = $1/[2\,t_{\rm univ}(z)]$, where $t_{\rm univ}(z)$ is the age of the Universe \citep[e.g.,][]{Tacchella2022_Fast_slow_quenching, Park2022}. We identify 50 quiescent central galaxies at $z=1.5$ and measure their formation timescales ($t_{50}^{90}$). The left panel of Fig.~\ref{fig:sfh_tng100_rapid_formation} shows the formation timescales ($t_{50}^{90}$) of all quiescent galaxies at $z\sim1.5$ as a function of their quenching epoch, $z_{\rm quench}$, defined as the epoch when their SFR starts to drop below sSFR = $1/[2\,t_{\rm univ}(z)]$.

In Fig.~\ref{fig:number_density2}, we have estimated that $\sim4\%$ of the growth of quiescent galaxies (i.e., 4\% of the newly quenched galaxies) at $z\sim1.5$ can be explained by the transition of the rapidly quenched galaxies. The left panel of Fig.~\ref{fig:sfh_tng100_rapid_formation} shows that there are only $\sim 5-6$ galaxies in TNG100 that are quenched at $z\sim1.5$, and only 4\% of them can be rapidly quenched analogs according to our estimate. Indeed, these TNG100 galaxies quenched at $z\sim1.5$ have much longer formation timescales. The formation timescales (16th-84th) of our 12 UVISTA rapidly quenched galaxies at $z\sim1.5$ are shown as a blue hatched box for comparison.

Since there are too few rapidly quenched analogs quenched at $z\sim1.5$, we instead select the 12 quiescent galaxies (regardless of quenching epochs) that have the shortest formation timescales; these are all quenched at earlier epochs. The 12 selected galaxies are marked with pentagons in the left panel and their SFHs are shown in the right panels. The green solid line shows the SFH derived from the age distribution of all stellar particles within $3\,R_{\rm eff}$, and the hatched grey histogram shows the SFH of stellar particles formed ex situ and later accreted. The formation timescales are shown as orange horizontal bars. The quenching epoch $z_{\rm quench}$ and the mass fraction of the stellar particles formed ex situ $f_{ex\,situ}$ are given in the right corner of each panel.

Clearly, in 5 of the 12 galaxies, (possibly gas-rich) major mergers seem to have triggered the starbursts. They have star formation peaks shortly after the accretion of stars from other galaxies. 
On the other hand, the rest did not have major mergers, but still had starbursts and rapid quenching. 
This result is consistent with the results of \cite{Pathak2021} where they found that in TNG approximately half of the young quiescent galaxies at $z=2$ have significant mergers prior to $z=2$. They further found that those that had significant mergers have centrally concentrated young stellar populations, while the age gradient is rather flat for the young quiescent galaxies without mergers.

In all cases, the rapid quenching of massive quiescent TNG100 galaxies seems to be driven by kinetic AGN feedback. Many studies have shown that kinetic AGN feedback is responsible for quenching massive galaxies in the TNG model \citep[e.g.,][]{Weinberger2018}, while energy injected via thermal AGN mode is mostly radiated away, partially due to limited numerical resolution, thus inefficient at quenching. We measure the epoch when the kinetic AGN mode is turned on for each galaxy and mark it as a magenta arrow in each SFH panel, and we can see clearly that the quenching epochs are tightly related to the epochs when the kinetic AGN mode is turned on.

\section{Discussion}
\label{sec:discussion}

\subsection{Size evolution}
Galactic morphology and sizes are closely linked to the star formation activities of galaxies. In particular, star-forming and quiescent galaxies have distinct sizes at fixed mass, as found in many studies \citep[e.g.,][]{VanDerWel20143D-HST+CANDELS:3}; at lower mass ($\log(M_{\rm stellar}/M_\odot)<10.5$), star-forming galaxies tend to be larger with their extended star-forming disks, while quiescent galaxies likely develop centrally-concentrated bulge-dominated structures, resulting in compact sizes. Therefore, the size of galaxies is the reflection of how the structures have been transformed as galaxies evolve to quiescence, holding an important clue of the physical mechanisms behind quenching.

The sizes of the newly quenched galaxies are expected to depend on their formation channel \citep[e.g.,][]{Wu2018}. If the quenching mechanisms do not invoke significant structural changes, newly quenched (young quiescent) galaxies are expected to be similar or even larger as they are transformed from large star-forming progenitors (larger at later times). The possible quenching mechanisms include thermal AGN feedback that could heat the surrounding medium \citep[e.g.,][]{Croton2006, Somerville2008} or very low SF inefficiency due to a bulge \citep[e.g.,][]{Martig2009} or bar structures \citep[e.g.,][]{Khoperskov2018}. The addition of large newly quenched galaxies into the quiescent populations can explain the size growth of quiescent population with time. In this scenario, a clear size-age relation for quiescent galaxies is expected \citep[e.g.,][]{vanderWel2009, Carollo2013, Fagioli2016}.
Dry minor mergers can complicate this picture, as they can also increase the size of galaxies by adding accreted stars to the outskirts of galaxies \citep[e.g.,][]{vanDokkum2005, Naab2007}.
On the other hand, if the quenching mechanisms involve dissipative processes that funnels gas into the central region and triggers a central starburst, this would result in smaller sizes than existing (old) quiescent populations. Several studies have also shown that these quenched galaxies after starbursts develop high central stellar mass densities \citep[e.g.,][]{Tacchella2016b, Barro2017, Mosleh2017}.

In Fig.~\ref{fig:size_mass}, we have shown the size-mass relation for 6 of our 12 UVISTA rapidly quenched galaxies, and most of them, except for one galaxy, have smaller sizes than the size-mass relation found at their respective redshifts. Our results are consistent with previous studies where they found young quiescent galaxies (or PSBs) are more compact that older/normal quiescent galaxies \citep[e.g.,][]{Whitaker2012compact_PSB, Belli2015, Almaini2017, Maltby2018, Wu2018}. This supports the scenario in which significant structural changes, particularly compaction triggered by central starbursts, precede quenching for young quiescent galaxies, most commonly found at high redshifts.

\subsection{Physical mechanisms driving the starburst and rapid quenching phases}
What are the physical mechanisms that quench galaxies rapidly while making them compact? One important finding from this work is that they are not just rapidly quenched but also rapidly formed. As we have seen in Fig.~\ref{fig:SFH_UVISTA_phot_only}, all of our 12 UVISTA rapidly quenched candidates had intense starbursts a few hundred Myrs before the observations, followed by rapid quenching. Their compact sizes (even more compact than normal quiescent galaxies, shown in Fig.~\ref{fig:size_mass}) also supports the idea that gas flows into the central region, triggering starbursts in this limited region. At high redshifts, the compaction processes appear to be more common, and the possible mechanisms include mergers, violent disk instability or misaligned gas stream \citep[e.g.,][]{Zolotov2015CompactionNuggets, Tacchella2016b}.

It is then reasonable to find the galactic gas being depleted after the starburst, but temporarily, as galaxies would be replenished with newly cooled gas from hot gas reservoirs. More sustainable feedback is required to keep post-starburst galaxies quiescent. Many quenching mechanism has been proposed; of these, mechanisms involving gas removal are thought to be rapid quenching process \citep[e.g.,][]{Man2018, Trussler2020} - for example, ejective AGN feedback \citep[e.g.,][]{Silk1998} or ram pressure stripping \citep[e.g.,][]{Gunn1972} (commonly found in the cluster environment). Since at high redshifts environmental diversity is not as dramatic as in the local universe, and the ram pressure stripping would be more effective for smaller satellite with weaker restoring force, the ejective AGN feedback is a more likely scenario for the massive post-starbursts explored in this study.

Indeed, we have detected AGN-driven emission line characteristics for two of the three galaxies with spectral features detected (See Section~\ref{sec:emission_lines}). The kinematics of the broad components, $\rm FWHM >1000\,km\,s^{-1}$, suggest that these outflows are gravitationally unbound and based on the very low level of H$\alpha$ emission, it is also very unlikely that the outflows are driven by SF. Also, in the TNG100 simulation, the rapid quenching seems to be caused by the kinetic AGN feedback, as explored in many previous studies \citep[e.g.,][]{Weinberger2018, Nelson2019, Luo2020WhatGalaxies, Terrazas2020}; the quenching epochs of our rapidly quenched analogs are tightly related to the epoch when the kinetic AGN mode is turned on (See Fig.~\ref{fig:sfh_tng100_rapid_formation}).

Then, the question remains: what initiates the kinetic AGN feedback? Clearly, 5 of 12 simulated rapid quenched analogs we explored in Fig.~\ref{fig:sfh_tng100_rapid_formation} have (possibly gas-rich) major mergers ($f_{ex\,situ}>0.25$) that trigger the starburst. The mass of the SMBH could have also significantly increased as a result of major mergers, releasing high amount of kinetic energy to the surrounding medium and eventually leading to rapid quenching. On the other hand, for the rest of the galaxies in Fig.~\ref{fig:sfh_tng100_rapid_formation} (especially galaxies that are quenched at higher redshifts, $z_{\rm quench}>3$), starbursts seem to be triggered by other mechanisms not involving major mergers, such as minor mergers or high rates of gas inflow. The gas inflow to the central regions that led to the starbursts could have fueled the SMBH and initiated the AGN feedback \citep[e.g.,][]{Tacchella2016aThereplenishment}.

In summary, the rapidly quenched galaxies at $z\sim1.5$ are also rapidly formed with intense starbursts a few hundred Myrs before quenching. The rapid quenching might be driven by the kinetic AGN feedback, as suggested by observed emission-line characteristics (high [NII]/H$\alpha$ and the kinematics of broad emission components) and based on the quenching of TNG100 simulated galaxies. However, the dominant mechanism for the starburst is not clear. We have found some cases in which major mergers trigger the starburst, but in other cases, starbursts seem to require other mechanisms.

\subsection{Overall picture of galaxy quenching at high redshift}
As discussed in Section~\ref{sec:the_role_of_rapid_quenching}, the rapid quenching phase is a very short transition phase with a timescale of $\approx 300\,\rm Myr$, and only a small fraction of quiescent galaxies ($4\%$ at $z\sim 1.4$ and $\sim10\%$ at $z\sim1.8$, see Fig.~\ref{fig:number_density2}) seem to have gone through this rapid quenching phase. Then, the question remains: what is the overall picture of galaxy quenching at high redshifts? To answer this question, here we explore the SFH of galaxies in other parts of the $UVJ$ diagram. Moreover, by exploring galaxies outside our main selection we check whether there is a substantial population of rapidly quenched galaxies we have missed because of high dust attenuation.

\begin{figure*}
    \centering
    \includegraphics[width=\textwidth]{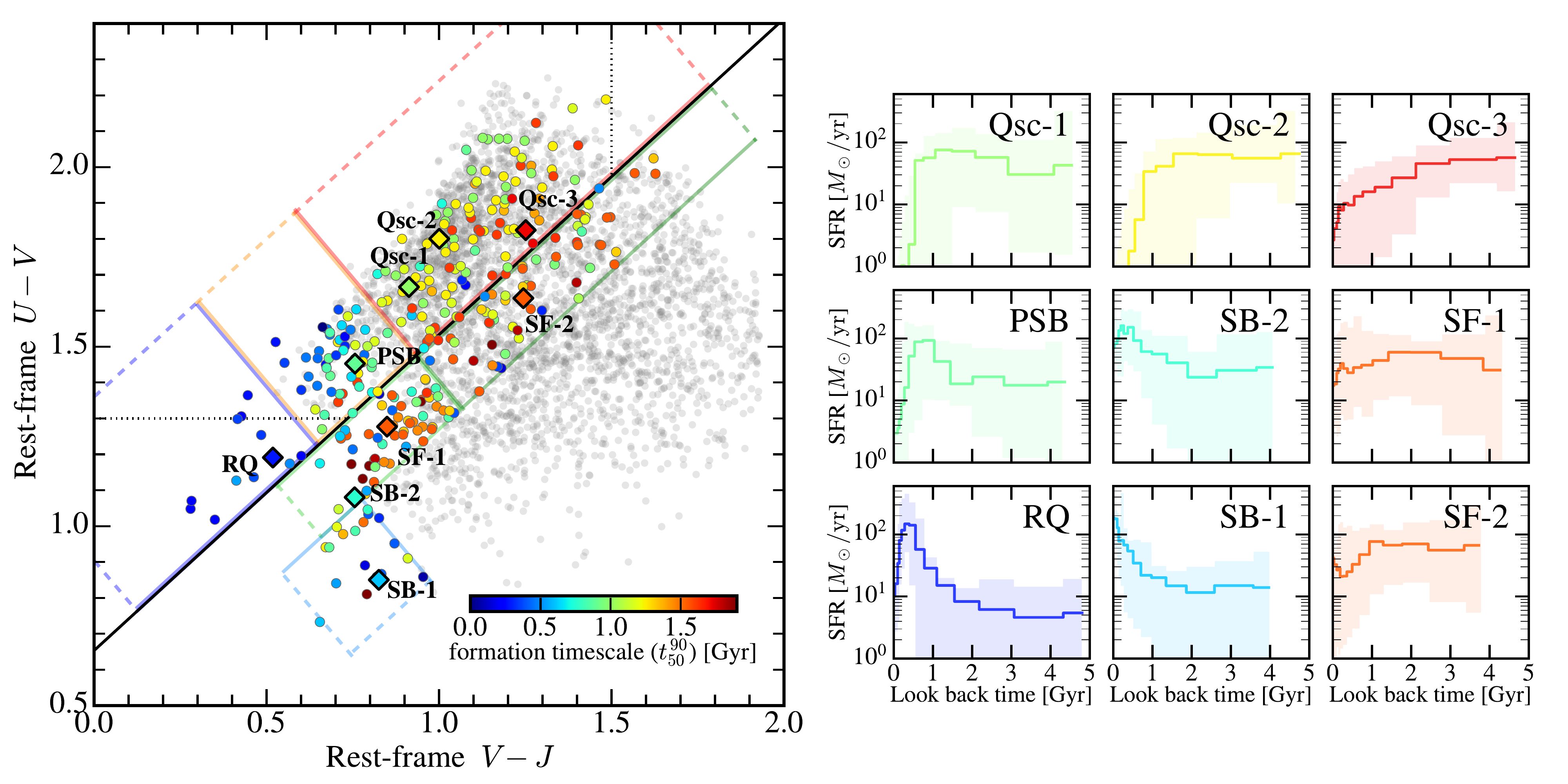}
    \caption{Formation timescale and SFH of UVISTA galaxies in different parts of the $UVJ$ diagram. (Left) UVISTA galaxies ($\log(M_{\rm stellar}/M_\odot)>10.6$ and $1.25<z<1.75$) in the $UVJ$ plane, color-coded by their formation timescales. The parent sample is shown as gray circles in the background. Note that the number of galaxies where formation timescales are measured (thus, color-coded) in each selected region is arbitrary, for the purpose of finding a qualitative trend. (Right) The SFH of example galaxies in different regions (highlighted as diamonds in the left panel) derived from SED fitting on photometry using Prospector. The MAP distribution is shown in a solid line with shades including 95\% posterior distribution. We show that galaxies in different parts of the $UVJ$ plane have different SFH, and thus, different formation timescales.}
    \label{fig:UVJ_SFH_all}
\end{figure*}

We measure the formation timescale of UVISTA galaxies in different regions of the $UVJ$ diagram (see the selected regions shown as boxes of different colors in Fig.~\ref{fig:UVJ_SFH_all}). Galaxies in Fig.~\ref{fig:UVJ_SFH_all} are color-coded by their formation timescale $t_{50}^{90}$. The parent sample is shown as gray circles in the background. Note that the number of galaxies where formation timescales are measured (thus, color-coded) in each selected region is arbitrary, for the purpose of finding a qualitative trend, so that the density distribution of colored points does not represent the density distribution of the parent sample. The right panels of Fig.~\ref{fig:UVJ_SFH_all} shows the SFH of example galaxies in different regions (highlighted as diamonds in the left panel), derived from SED fitting on photometry using Prospector. The MAP distribution is shown as a solid line with shades including 95\% posterior distribution, and the color of the lines represents the formation timescale of the particular galaxy.

It is clear that galaxies in different parts of the $UVJ$ diagram have different formation timescales and star formation histories. Galaxies in the rapid quenching region (as an example galaxy ``RQ'' shown on the right panel) are the ones that are the most rapidly formed (by very intense starbursts followed by rapid quenching) with the formation timescales typically less than 0.5 Gyr. Galaxies in the PSB region are a mixture of galaxies similar to rapidly quenched galaxies and others with less bursty SFH (an example is galaxy ``PSB''). Quiescent galaxies have different quenching histories depending on their location in the $UVJ$ diagram, including galaxies that are rapidly quenched without having starburst, thus, having rather flat SFH in the past (as the galaxy ``Qsc-1'' and ``Qsc-2''), and galaxies that are slowly quenched (for example galaxy ``Qsc-3'').

The majority of galaxies with the bluest $U-V$ colors (in a sky-blue box region) have SFHs that are rapidly rising (galaxy ``SB-1''), and could be the progenitors of rapidly quenched galaxies if their star formation is abruptly terminated in the near future. Normal SF galaxies have a rather flat SFH (like galaxy ``SF-1'' or ``SF-2''), and if they experience bursts in star formation they may evolve to the bluer side of the $UVJ$ diagram (as galaxy ``SB-1'' or ``SB-2''). On the other hand, if they are quenched, they will evolve into the quiescent region as galaxy ``Qsc-1'', if quenched rapidly, and as galaxy ``Qsc-3'', if quenched slowly (see Appendix~\ref{appendix} for the recovery of rapid quenching in older galaxies).

A few galaxies in the quiescent and star-forming region of the $UVJ$ diagram appear to have short formation timescales (colored as blue points in Fig.~\ref{fig:UVJ_SFH_all}). These are rapidly quenched galaxies but dust-rich so that they have much redder $UVJ$ colors than the rapidly quenched galaxies we have selected. The existence of these dusty rapidly quenched galaxies will likely increase our estimated fractions of the quiescent galaxies explained by the transition from rapid quenching in Fig.~\ref{fig:number_density2}, but not by a substantial amount.

\begin{figure*}
    \centering
    \includegraphics[width=\textwidth]{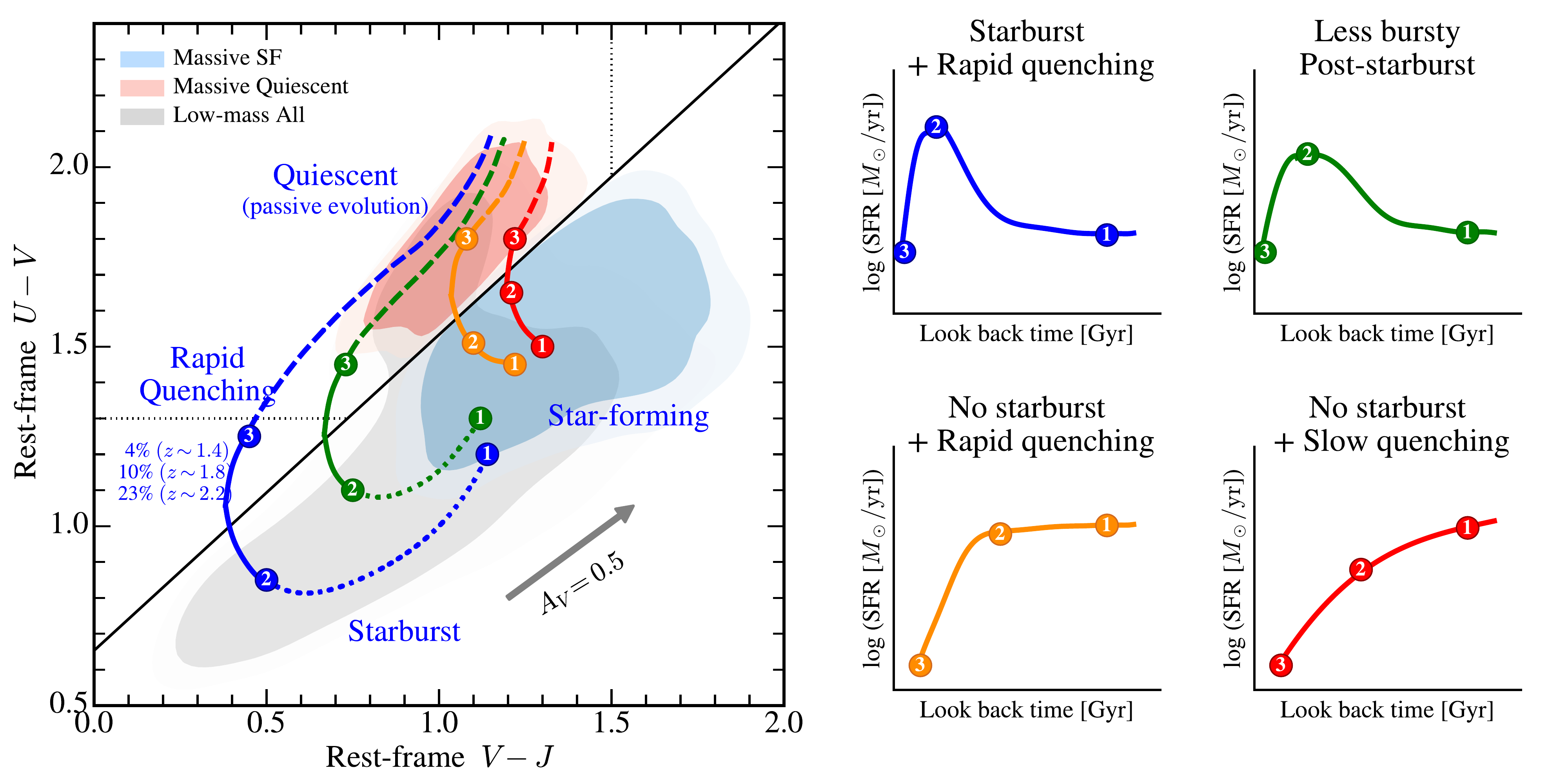}
    \caption{Schematic diagram summarizing different evolutionary tracks from star-forming to quiescence in the $UVJ$ diagram depending on the SFHs (Blue: Galaxies that have starburst followed by rapidly quenching. Green: Galaxies having less bursty SFH. Orange: Galaxies that do not have starburst but simply have rapid quenching. Red: Galaxies that are slowly quenched without starburst). The number in each colored circle in the $UVJ$ space corresponds to the stage in SFH of that number in the right panel. The tracks for passive evolution after quenching are plotted in dashed lines. The blue and red contours represent the distribution of massive ($\log(M_{\rm stellar}/M_\odot)>10.6$) star-forming ($C_Q < 0.49$) and quiescent ($C_Q > 0.49$) galaxies, and the gray contour shows the distribution of lower-mass ($9.8<\log(M_{\rm stellar}/M_\odot)<10.6$) galaxies. }
    \label{fig:UVJ_summary}
\end{figure*}

Fig.~\ref{fig:UVJ_summary} is a schematic diagram summarizing different evolutionary tracks from star-forming to quiescence in the $UVJ$ diagram depending on the SFHs. We show the $UVJ$ color evolution of galaxies that have starburst followed by rapid quenching (blue), galaxies having less bursty SFH (green), galaxies that do not have starburst but simply have rapid (orange) and slow (red) quenching. For each case, the number in each colored circle in the $UVJ$ space corresponds to the stage in SFH of that number in the right panel. The blue and red contour represents the region where it includes $1\sigma$ and $2\sigma$ of massive ($\log(M_{\rm stellar}/M_\odot)>10.6$) star-forming ($C_Q < 0.49$) and quiescent ($C_Q > 0.49$) galaxies, respectively.

For galaxies that had starbursts in the past, the stellar mass would have also increased significantly during the starburst. Their star-forming progenitors before starburst (numbered ``1''), therefore, have much smaller masses. We plot the distribution of slightly lower-mass galaxies ($9.8<\log(M_{\rm stellar}/M_\odot)<10.6$) in gray contour, and generally, lower-mass star-forming galaxies have much bluer colors. It is not clear how dusty these low-mass star-forming progenitors were, i.e., where they were located in the $UVJ$ space, so we mark this part of the evolutionary tracks in dotted lines. As galaxies go through starburst triggered by mergers and/or disk instability and become compact, they would evolve to the bluer corner of $UVJ$ diagram (numbered ``2''). They would then evolve to the rapid quenching region as they are quenched (numbered ``3''). The passive evolution after quenching (the track beyond the point ``3'') is plotted in dashed lines. If they have less extreme starburst, they get only slightly bluer and then quenched, as shown in the green track. Thus, only the galaxies that have extremely bursty SF followed by rapid quenching would end up in our rapid quenching region: the fraction of quiescent galaxies that would go through this extreme phase increases with redshift (see Section~\ref{sec:number_density}): 4\% at $z\sim1.4$, 10\% at $z\sim1.8$, and 23\% at $z\sim2.2$. Once they are quenched, they would passively evolve to the quiescence sequenced shown in red contour (dashed tracks).

Galaxies that do not experience starburst would not evolve to bluer side of the $UVJ$ diagram, but their color would just get redder, as shown in the orange (rapid quenching) and red (slow quenching) tracks. Because they are quenched without a starburst, their mass would not have increased much compared to before quenching. Thus, their star-forming progenitors before quenching most likely reside in the massive star-forming galaxy sequence, represented as a blue contour.

The evolution of the amount of dust has not been taken into account in this schematic diagram. Broadly, dust would move galaxies to the diagonal direction in the $UVJ$ diagram, following the arrow, assuming the \citet{Calzetti2000} extinction law. As galaxies go through starbursts and rapid quenching, the amount of dust changes as well, but the detailed evolution is not entirely clear. There have been studies suggesting that the wide range of dust across the galaxies on the $UVJ$ space could be entirely the effect of galaxy inclination \citep[e.g.,][]{Zuckerman2021}, which complicates the picture even more since it links color evolution and  morphological transformation. More detailed studies of the dust evolution during starburst and quenching and how this evolution is related to inclination and morphology are needed for future work. 

\section{Summary and Conclusion}
\label{sec:summary_conclusion}

In this work, we explored 12 young ($<300\,\rm Myr$) massive ($\log(M_{\rm stellar}/M_\odot)>10.6$) quiescent galaxies at $z\sim1.5$, selected based on their location in the $UVJ$ diagram, as rapidly quenched candidates. Here we summarize our results.

\begin{enumerate}
    \item From SED fitting of their photometric data using \texttt{Prospector}, we confirm that our young quiescent sample had intense starbursts in the past and then rapidly quenched (thus, they are truly ``post-starburst''). They all formed very rapidly with formation timescale (the time it takes to form 50\% to 90\% of their final stellar mass) of $t_{50}^{90}\approx 320\,\rm Myr$ (Fig.~\ref{fig:SFH_UVISTA_phot_only}). Using the 3D-DASH data, we find that most of the sample has compact sizes, even smaller than normal quiescent galaxies (Fig.~\ref{fig:size_mass}). Their compact sizes provide another piece of evidence that they had central starbursts in the past.
    
    \item We performed Magellan/FIRE spectroscopic observations of seven galaxies, and we confirm that they are truly quenched without H$\alpha$ emission. We detected absorption lines for two of the most massive galaxies in our sample (UVISTA 169610 and 174150) and found that their quenching history is slightly better constrained with the spectroscopic data, but photometric data seems to be sufficient to provide an estimate of how rapidly galaxies are quenched. Of the three galaxies with spectral features, two of them show signs of AGN activity with high [NII]/H$\alpha$ ratios and broad emission components (FWHM $>1000\,\rm km\,s^{-1}$). The other galaxy, the reddest galaxy among our sample, does not show any emission features, suggesting that ionized gas may have already been blown away by AGN activity.
    
    \item From the star formation histories  of the 12 UVISTA galaxies we infer the time it takes to cross the rapid quenching region, finding a transition time of $\approx 300 \rm\, Myr$. Using this transition time, we calculate how much of the growth rate of quiescent galaxies can be explained by the transition from rapid quenching and found that the rapid quenching phase accounts for only a small fraction of the growth of the quiescent population at $z\sim1.5$. However, the fraction of quiescent galaxies that have gone through the rapid quenching phase increases with redshift: 4\% at $z\sim1.4$, 10\% at $z\sim1.8$, and 23\% at $z\sim2.2$ (Fig.~\ref{fig:number_density2}).

    \item We identified 50 massive quiescent galaxies ($10.6<\log(M_{\rm stellar}/M_\odot)<12.0$) at $z\sim1.5$ in the TNG100 simulation and measured their formation timescales. We found that galaxies quenched earlier are formed more rapidly and select 12 galaxies that have shortest formation timescales. In 5 of these 12 galaxies, starbursts appear to have started shortly after major mergers, while the others did not have major mergers, but still had starbursts and rapid quenching. In all cases, the rapid quenching in TNG100 is driven by the kinetic AGN feedback. 
    
    \item We studied the SFH of galaxies in other parts of the $UVJ$ diagram from \texttt{Prospector} fitting on their photometric data and found that their formation timescales and histories depend on the location in the $UVJ$ diagram. We show how galaxies would evolve in the $UVJ$ space when they have/do not have starburst and when they are rapidly/slowly quenched (Fig.~\ref{fig:UVJ_summary}).  
    
\end{enumerate}

The mechanisms by which high-redshift galaxies in the early universe stop forming stars and become quiescent remain a puzzle. Based on our study of the youngest quiescent galaxies at $z\sim1.5$, we conclude that these massive quiescent galaxies quenched at high redshifts are \textit{not just rapidly quenched but also rapidly formed with a starburst}. The starburst appears to have occurred in the central regions given their compact sizes, likely triggered by dissipative processes such as gas-rich mergers, migration of star-forming clumps, or accretion of gas streams. In the TNG simulation, rapid quenching is driven by SMBH feedback that have grown as a result of the processes that led to central starbursts prior to rapid quenching.

The importance of rapid quenching becomes more significant at higher redshifts, as we estimated in this work. In particular, at high redshift ($z>3$), where the age of the Universe is less than 2 Gyr, all of the quiescent galaxies are expected to be quenched rapidly. We expect that future observations using, e.g., JWST, of the first quiescent galaxies in the early Universe will shed more light on our understanding of the rapid quenching and further galaxy evolution. Also, future simulations that use realistic AGN feedback models and have sufficient resolution to resolve AGN jets and central starbursts will help us understand the detailed sequential process of the rapid formation at high redshifts.

\begin{acknowledgments}
We thank Dylan Nelson and Annalisa Pillepich for useful discussions. S.B. is supported by the Italian Ministry for Universities and Research through the \emph{Rita Levi Montalcini} program. C.C. acknowledges support from NSF AST-1908748. S.C. wishes to acknowledge funding under HST-GO-16259. R.E. acknowledges the support from the Institute for Theory and Computation at the Center for Astrophysics as well as grant numbers 21-atp21-0077, NSF AST-1816420 and HST-GO-16173.001-A for very generous support.

\end{acknowledgments}

\appendix
\renewcommand{\thefigure}{A\arabic{figure}}
\setcounter{figure}{0}

\section{How does the derived SFH of rapidly quenched galaxies change after passive evolution?}
\label{appendix}

Fig.~\ref{fig:UVJ_SFH_all} shows the continuous distribution of quiescent galaxies with a clear age trend along the diagonal direction: from youngest quiescent galaxies that are rapidly and recently quenched to old quiescent galaxies. 
Once galaxies are quenched, they will passively evolve along the diagonal direction of the $UVJ$ diagram. Thus, the old quiescent population consists of galaxies that have passively evolved after rapid quenching in addition to galaxies that are slowly quenched. In Section~\ref{sec:number_density}, we calculated how much of the growth rate of the quiescent population can be explained by the transition from rapid quenching and found that only a small fraction of quiescent galaxies appear to have gone through the rapid quenching phase ($\approx4\%$ at $z\sim1.4$). We aim to confirm this finding by checking directly how the SFH from \texttt{Prospector} fitting would change as galaxies passively evolve. In this section, we test what the SFH of rapidly quenched galaxies would look like if they were observed a few Gyrs later and compare it with the SFH of old quiescent galaxies.

\begin{figure*}
    \centering
    \includegraphics[width=\textwidth]{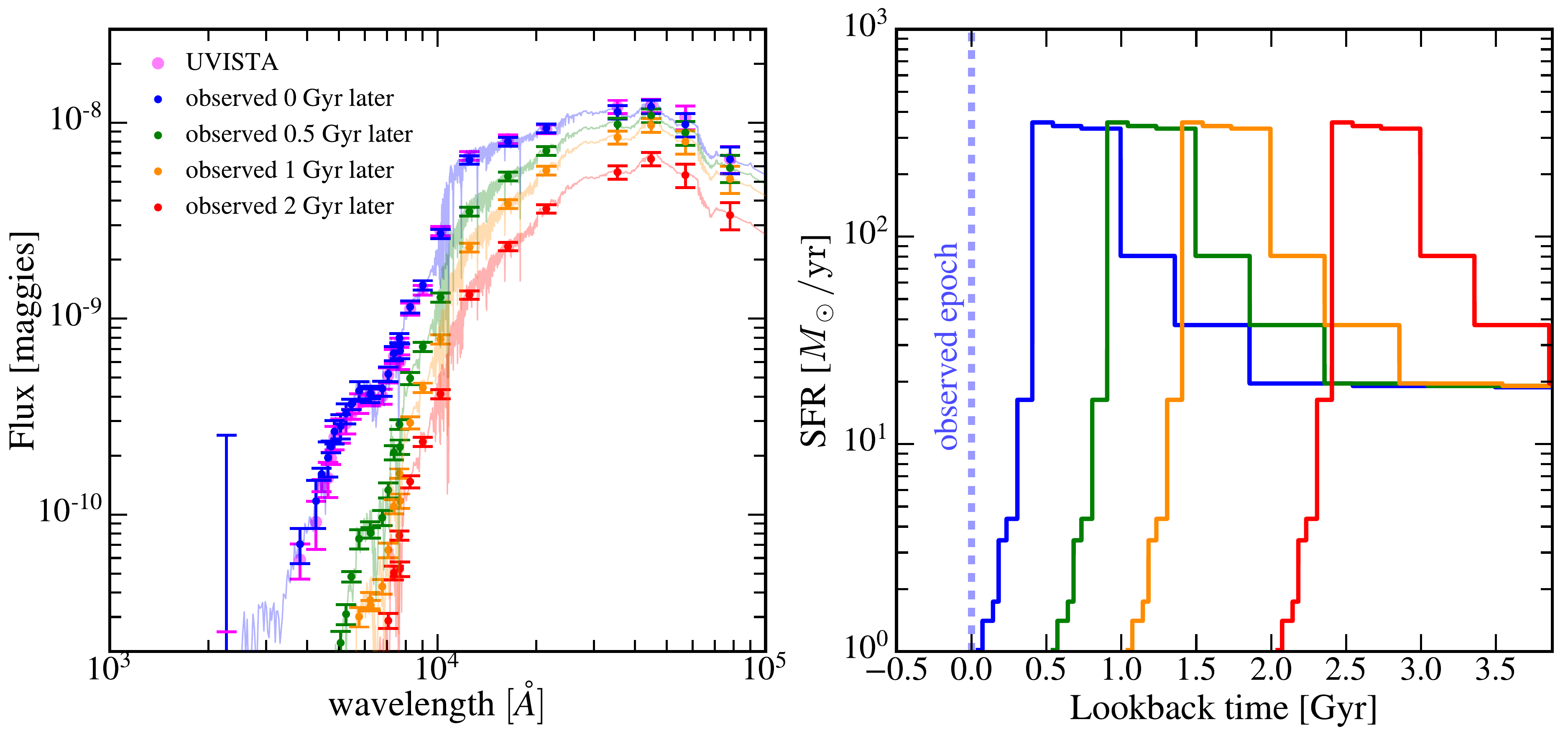}
    \caption{(Left) Mock photometry of UVISTA 174150 generated with MAP SFH pushed back by 0 Gyr (blue), 0.5 Gyr (green), 1 Gyr (orange), and 2 Gyr (red). Each colored line shows the mock SED generated with each SFH. We assume the same signal-to-noise ratio of the photometric measurements for each filter used in the UVISTA survey and generate mock noise shown as error bars. The magenta points are the photometric data provided by the UVISTA catalog. (Right) MAP SFH of UVISTA 174150 pushed back to account for passive evolution (same color code used as the left panel). The oldest bins are removed to assume the same observed epoch/redshift. }
    \label{fig:mock_phot_UVISTA_174150}
\end{figure*}

\begin{figure*}
    \centering
    \includegraphics[width=\textwidth]{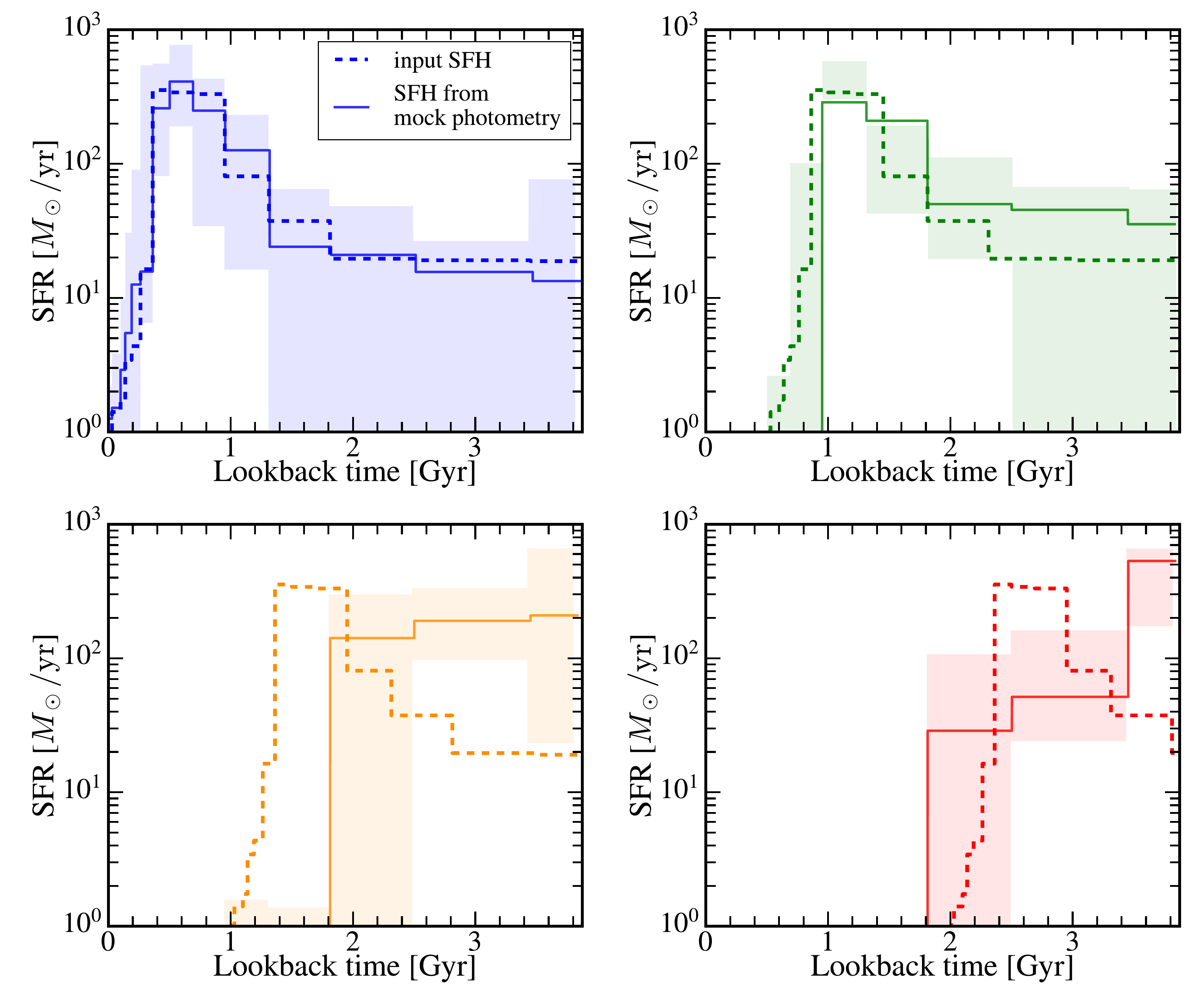}
    \caption{SFH from \texttt{Prospector} fitting on mock photometry generated with the SFH shifted back by 0 Gyr (blue), 0.5 Gyr (green), 1 Gyr (orange), and 2 Gyr (red). The dashed line shows the input SFH used to generated mock photometry, and the solid line is the MAP distribution from \texttt{Prospector} fit on the mock photometry. The shaded regions include 95\% of the posterior distribution.}
    \label{fig:SFH_of_mock_phot}
\end{figure*}

To take into account the passive evolution, we push back the SFH of a rapidly quenched galaxy and remove the oldest bins so that we can still assume that the galaxy is observed at the same redshift. The right panel of Fig.~\ref{fig:mock_phot_UVISTA_174150} shows the MAP SFH of UVISTA 174150 (blue) and the SFH pushed back by 0.5 Gyr (green), 1 Gyr (orange), and 2 Gyr (red). We then generate a stellar population using \texttt{FSPS} by feeding each SFH and calculate the magnitudes (flux) for the UVISTA photometric filters. We assume the same signal-to-noise ratio of the photometric measurements for each filter used in the UVISTA survey. The left panel of Fig.~\ref{fig:mock_phot_UVISTA_174150} shows the mock photometric data of UVISTA 174150 when observed 0 Gyr (blue), 0.5 Gyr (green), 1 Gyr (orange), 2 Gyr (red) later. The magenta points are the photometric data provided by the UVISTA catalog. We run \texttt{Prospector} on the mock photometric data generated with shifted SFHs. Fig.~\ref{fig:SFH_of_mock_phot} shows the SFH from \texttt{Prospector} fitting on mock photometry generated with each shifted SFH. The solid line shows the MAP distribution with shaded regions indicating 95\% of the posteriors. The dashed line represents the input SFH.

The starburst and rapid quenching feature does not seem to be well reproduced with \texttt{Prospector} when observed $>1\,\rm Gyr$ after quenching. This is probably because the time (age) bins are much coarser at older ages and it is more difficult to distinguish the relative contributions of older stellar populations. Still, even observed a few Gyrs later, a sharp cutoff in the SFH is reproduced by \texttt{Prospector} fitting. 
Among the random $\sim 100$ quiescent galaxies we present in Fig.~\ref{fig:UVJ_SFH_all}, only $\sim5-6$ galaxies seem to have sharp truncations in SFH a few Gyrs ago. This gives us another piece of evidence that only a small fraction of quiescent galaxies are the descendants of rapidly quenched galaxies.

\bibliography{references}{}
\bibliographystyle{aasjournal}
\end{document}